\documentclass[%
 amsmath,amssymb,
 reprint,%
superscriptaddress,
]{revtex4-1}

\usepackage{graphicx}
\usepackage{dcolumn}
\usepackage{bm}

\usepackage[utf8]{inputenc}
\usepackage[T1]{fontenc}
\DeclareMathAlphabet{\mathcal}{OMS}{cmsy}{m}{n}
\usepackage{mathptmx}

\begin{document}

\title{Mechanism of paramagnetic spin Seebeck effect}

\author{Koichi Oyanagi}
\email{k.0yanagi444@gmail.com}
\affiliation{Faculty of Science and Engineering, Iwate University, Morioka 020-8551, Japan}
\affiliation{Institute for Materials Research, Tohoku University, Sendai 980-8577, Japan}

\author{Saburo Takahashi}%
\affiliation{Institute for Materials Research, Tohoku University, Sendai 980-8577, Japan}
\affiliation{ WPI Advanced Institute for Materials Research, Tohoku University, Sendai 980-8577, Japan
}%

\author{Takashi Kikkawa}
\affiliation{Institute for Materials Research, Tohoku University, Sendai 980-8577, Japan}
  \affiliation{ WPI Advanced Institute for Materials Research, Tohoku University, Sendai 980-8577, Japan
}
\affiliation{ 
Department of Applied Physics, University of Tokyo, Tokyo 113-8656, Japan
}%

\author{Eiji Saitoh}
\affiliation{Institute for Materials Research, Tohoku University, Sendai 980-8577, Japan}
\affiliation{ WPI Advanced Institute for Materials Research, Tohoku University, Sendai 980-8577, Japan
}
\affiliation{ 
Department of Applied Physics, University of Tokyo, Tokyo 113-8656, Japan
}%
\affiliation{Institute of AI and Beyond, The University of Tokyo, Tokyo 113-8656, Japan
}%
\affiliation{ 
Advanced Science Research Center, Japan Atomic Energy Agency, Tokai 319-1195, Japan
}%

\date{\today}

\begin{abstract}

We have theoretically investigated the spin Seebeck effect (SSE) in a normal metal (NM)/paramagnetic insulator (PI) bilayer system. Through a linear response approach, we calculated the thermal spin pumping from PI to NM and backflow spin current from NM to PI, where the spin-flip scattering via the interfacial exchange coupling between conduction-electron spin in NM and localized spin in PI is taken into account. We found a finite spin current appears at the interface under the difference in the effective temperatures between spins in NM and PI, and its intensity increases by increasing the density of the localized spin $S$. Our model well reproduces the magnetic-field-induced reduction of the paramagnetic SSE in Pt/Gd$_3$Ga$_5$O$_{12}$ experimentally observed when the Zeeman energy is comparable to the thermal energy, which can be interpreted as the suppression of the interfacial spin-flip scattering. The present finding provides an insight into the mechanism of paramagnetic SSEs and the thermally induced spin-current generation in magnetic materials.

\end{abstract}

\maketitle

\section{INTRODUCTION}
Spin Seebeck effect (SSE) \cite{uchida2008observation, uchida2014longitudinalreview, uchida2016thermoelectric, uchida2021transport, kikkawa2022spin} refers to the spin-current generation from a temperature gradient applied to a magnet. The generated spin current flows along the temperature gradient and can be detected as a voltage signal in an attached normal metal (NM) electrode, such as Pt, using the inverse spin Hall effects (ISHEs) \cite{azevedo2005dc, saitoh2006conversion, kimura2007room, valenzuela2006direct, sinova2015spin}. Up to now, the SSEs have been investigated in various magnetically ordered materials, including ferrimagnets \cite{uchida2010spin, uchida2014longitudinal, geprags2016origin, oh2021scalable}, ferromagnets \cite{ito2019spin, mallick2019role}, and antiferromagnets \cite{seki2015thermal, wu2016antiferromagnetic, rezende2016theory, li2020spin, reitz2020spin, kikkawa2021observation}. In the mechanism of the SSEs in the ordered magnets, magnon excitation plays an important role. Xiao $et$ $al.$ formulated the thermal spin pumping theory for the SSEs in a NM/ferromagnetic insulator (FM) bilayer \cite{xiao2010theory}. In this mechanism, the temperature difference between the effective magnon temperature in FM and electron temperature in NM generated by the applied temperature gradient creates an imbalance between thermal spin pumping from FM to NM and the backflow spin current from NM to FM, resulting in a finite spin current across the interface \cite{bender2012electronic, adachi2013theory}. Subsequently, Rezende $et$ $al.$ proposed another mechanism of the SSEs originating from the bulk magnon transport induced by the temperature gradient based on diffusion equations of magnons \cite{rezende2014magnon, rezende2020fundamentals}.

The SSEs are also found in paramagnetic materials, where the conventional magnon excitation/transport can not be responsible for the mechanism of the SSEs. 
Wu $et$ $al.$ reported that the SSEs in paramagnetic insulators (PIs) Gd$_3$Ga$_5$O$_{12}$ and DyScO$_3$ at low temperatures close to their Curie-Weiss temperatures \cite{wu2015paramagnetic}. By measuring the SSEs across the phase transition temperatures, the paramagnetic SSEs have been found in the paramagnetic phase of CoCr$_2$O$_4$ \cite{aqeel2015spin}, FeF$_2$ \cite{li2019spin}, SrFeO$_{3}$ \cite{hong2019spin}, and CrSiTe$_{3}$ and CrGeTe$_{3}$ \cite{ito2019spin}.
The SSEs from paramagnets were also found in the one-dimensional (1D) quantum spin liquid (QSL) system Sr$_2$CuO$_3$ \cite{hirobe2017one, hirobe2018generation} and spin dimer systems CuGeO$_3$ \cite{chen2021triplon}, Pb$_{2}$V$_{3}$O$_{9}$ \cite{xing2022spin}, and VO$_{2}$ \cite{luo2022spin}. 
The SSEs in quantum spin systems are attributed to the thermal generation of exotic spin excitations in the systems, i.e., spinons in the 1D QSL and triplons in the spin dimer system, respectively. However, the mechanism of the thermally induced spin-current generation in a junction of NM and classical PI is not established; only a few theoretical studies addressed the spin currents in NM/PI junctions \cite{okamoto2016spin, yamamoto2019spin, zhang2019theory, nakata2021magnonic}. 

In this paper, we developed a theoretical model of the SSE in a bilayer of a NM and classical PI by considering the spin-flip scattering due to the interfacial exchange interaction between conduction-electron spins in NM and localized spins in PI.
We calculated thermal spin pumping from PI to NM and back-flow spin current from NM to PI at the NM/PI interface based on a linear response formalism \cite{kajiwara2010transmission, takahashi2010spin}.
The system is characterized by the effective temperatures of the localized spins in PI, $T_{\mathrm{PI}}$, and conduction electron spins in NM, $T_{\mathrm{NM}}$.
The finite spin current $J_{\mathrm{s}}$ arises when $T_{\mathrm{PI}} \neq T_{\mathrm{NM}}$ due to the imbalance between the thermal spin pumping and back-flow spin current, and its intensity is proportional to the density of the spin $S$ of the paramagnetic localized ion in PI. 
We describe $J_{\mathrm{s}}$ as a function of a single parameter $\xi \propto B/T$, where $B$ and $T$ represent the magnetic field $B$ and temperature $T$, respectively.
To compare the theoretical results with the experiments, we conducted the SSE measurements in a Pt/Gd$_3$Ga$_5$O$_{12}$(GGG) junction, where GGG is known as a classical paramagnet down to low temperatures. Our model well explains the experimental $B$ dependence of the paramagnetic SSE in Pt/GGG, which clarifies that the $B$ dependence of the paramagnetic SSE is attributed to the competition between the $B$-induced spin alignment (increasing $J_{\mathrm{s}}$) and the Zeeman gap opening (decreasing $J_{\mathrm{s}}$) with increasing $B$. A difference in the experimental and theoretical $T$ dependences was found, which may be attributed to the effect of the thermal conductivity in GGG and interfacial Kapitza conductance.

\section{SPIN CURRENT AT NM/PI}
In this section, we formulate the interfacial spin current in a NM/PI junction by taking the interfacial spin-flip scattering into account and apply the result to the case of the SSE. 
Conduction electrons in NM and localized spins in PI are coupled by the interfacial exchange interaction, leading to the exchange of the spin angular momentum and energy via the spin-flip scattering. 
Our formulation shows that the interfacial spin current is proportional to a Brillouin function of spins in PI, the density of the spin $S$ of magnetic ions in PI, and the difference between the distribution function of the PI and NM sides.

\begin{figure}[t]
\includegraphics{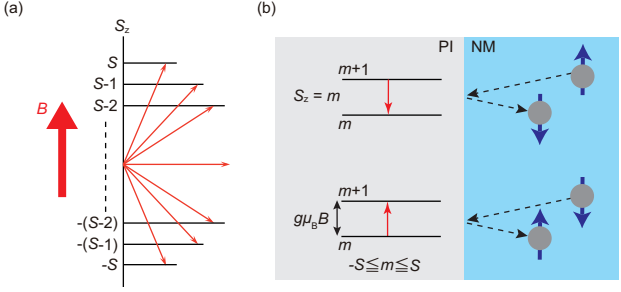}
\caption{\label{fig1} (a) A schematic illustration of Zeeman splitting of a spin-$S$ system. When the magnetic field $B$ is applied, the degeneracy is lifted to create $2S+1$ energy levels with the energy gap $\Delta E = g\mu_B B$ (b) A schematic illustration of spin exchange at the NM/PI interface. Upper (lower) spin-flip scattering corresponds to the spin absorption (injection) process from NM (PI) to PI (NM). }
\end{figure}

\subsection{Model}
We model the spin and energy transfer by the spin-flip scattering across a NM/PI interface biased by an external driving force such as a temperature gradient.
Figure \ref{fig1}(a) shows the energy levels of paramagnetic spin under the magnetic field $B$. At zero field, all spins are degenerated in a single energy level. 
By applying $B$, the spin degeneracy is lifted to split into different energy levels ($2S+1$). 
Each energy level is separated with the Zeeman energy of $g\mu_{\mathrm{B}} B$, where $g$ is the $g$-factor and $\mu_{\mathrm{B}}$ is the Bohr magneton.
We show a schematic illustration of the spin-flip scattering in Fig. \ref{fig1}(b). 
The spin-flip scattering causes the exchange of spin angular momentum of $\pm\hbar$ and energy of $\pm g\mu_{\mathrm{B}} B$ between a conduction electron in NM and a localized spin in PI, where $\hbar$ is the Dirac's constant. When an up-spin (down-spin) electron in NM interacts with a localized spin in PI, the spin-flip scattering lowers (raises) the energy state of the spin at the interface, generating a nonequilibrium spin state in PI [see the upper (lower) part of Fig. \ref{fig1}(b)].
We assume the local thermal equilibrium, which indicates that a state of the spin system X (X = NM or PI) is characterized by the effective temperature of $T_{\mathrm{X}}$ and spin chemical potential $\mu_{\mathrm{X}}$. Here, the spin chemical potential in PI, $\mu_{\mathrm{PI}}$, is similar to the magnon chemical potential in magnetic insulators \cite{cornelissen2016magnon}, but the physical picture is quite different because PI has no (well-defined) magnon. The magnon chemical potential represents the number of the nonequilibrium magnons, while $\mu_{\mathrm{PI}}$ corresponds to the amount of the nonequilibrium spins, i.e., the number of flipped spins different from the thermal equilibrium as a result of the spin-flip scattering at the interface.

 \begin{figure}[t]
\includegraphics{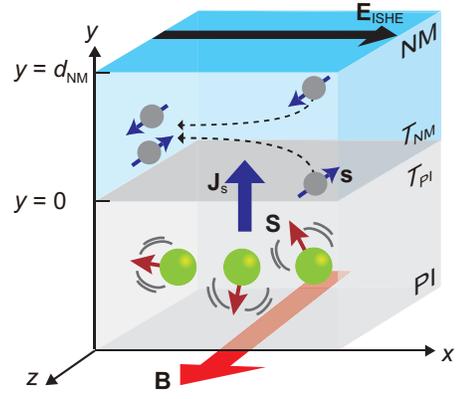}
\caption{\label{fig2-2} A schematic illustration of the NM/PI junction system. $\bold{B}, \bold{S}, \bold{s}, \bold{J}_\mathrm{s}$ and $\bold{E}_{\mathrm{ISHE}}$ indicate the applied magnetic field, spin of the localized magnetic ion in PI, spin of the conduction electron, interfacial spin current, and electric field due to the ISHE in NM, respectively.}
\end{figure}

Figure \ref{fig2-2} shows the considered NM/PI junction with the static magnetic field $\bold{B}$ in the $z$ direction. In the case of the SSE, when we apply the temperature difference between NM and PI in the $y$ direction ($T_{\mathrm{NM}} \neq T_{\mathrm{PI}}$), the net spin and energy exchange appears via the spin-flip scattering between the conduction electron spin $\bold{s}$ and localized spin $\bold{S}$ at the interface resulting in the generation of a spin current $\bold{J}_\mathrm{s}$ in the $y$ direction. The injected $\bold{J}_\mathrm{s}$ is converted into an electric field $\bold{E_{\mathrm{ISHE}}}$ via ISHE in NM and can be measured as a voltage signal $V_{\mathrm{ISHE}}$ in the $x$ direction.

 \subsection{Thermal average of spin}
First of all, we show the field-induced magnetization $\langle m\rangle$ in PI, corresponding to the thermal average of the magnetization component parallel to $\bold{B}$. The partition function of paramagnetic spins $Z = \Sigma^{S}_{m=-S}\mathrm{exp}(-g\mu_{\mathrm{B}}B/k_{\mathrm{B}}T)$, where $k_{\mathrm{B}}$ is the Boltzmann constant and $T$ is the temperature, gives $\langle m\rangle$ and the spin-spin correlation function $\langle m^2 \rangle$ as

\begin{align}
\langle m\rangle &=  \frac{2S+1}{2} \mathrm{coth} \left( \frac{2S+1}{2} \xi \right) - \frac{1}{2} \mathrm{coth}\left( \frac{1}{2} \xi \right) \notag \\
 &= SB_S(\xi), \label{Bf}\\
\langle m^2 \rangle &= S(S+1) - \langle m\rangle \mathrm{coth}\left( \frac{\xi}{2} \right) \label{Bf2},
\end{align}
where $B_{S}(\xi)$ is the Brillouin function of spin $S$ as a function of $\xi$, $\xi = C_1 B/T$ is the dimensionless ratio of the Zeeman energy to the thermal energy, and $C_1 = g\mu_{\mathrm{B}}/k_{\mathrm{B}}$. 
As the Brillouin function is parameterized by $\xi$ consisting of two independent parameters $B$ and $T$, an increase (decrease) of $\xi$ corresponds to the increased (decreased) $B$ at the fixed $T$ or the decreased (increased) $T$ at the fixed $B$ in actual experiment. 

\begin{figure}
\includegraphics{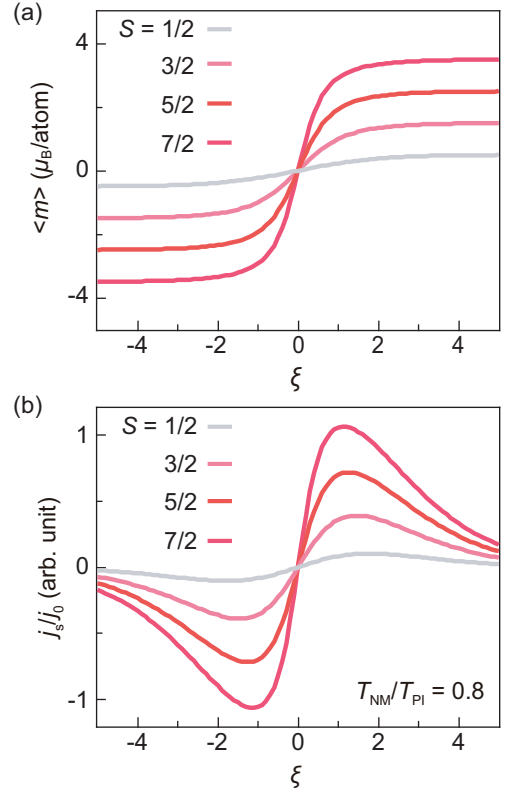}
\caption{\label{fig2c} The field-induced magnetization $\langle m\rangle$ (a) and normalized interfacial spin current density $j_{\mathrm{s}}/j^{0}_{\mathrm{s}}$ (b) with various $S$ as a function of $\xi$. The results in (b) are obtained with $T_{\mathrm{NM}}/T_{\mathrm{PI}} = 0.8$.}
\end{figure}

Figure \ref{fig2c}(a) shows the calculation of Eq. (\ref{Bf}) with different $S$ values. In a small $\xi$ regime (small $B$ or high $T$ condition), $\langle m\rangle$ linearly increases, showing a typical Curie-Weiss behavior. Above $\xi >1$, $\langle m\rangle$ is saturated to the value of $S$ because all spins align along the direction of $B$. 

We simply considered the non-interacting paramagnetic spins, i.e., the exchange interaction among the paramagnetic spins is zero.
This assumption is a good approximation when the system is under high $T$ or small $B$. 
However, finite exchange interaction in realistic materials becomes important when the system is under low $T$ or large $B$, where the energy scale of the system is comparable to that of the exchange interaction, characterized by the Curie-Weiss temperature. 
In comparison with experimental results (Sec. \ref{compari}), we take the actual exchange interaction into account using the lowest-order approximation of a molecular field theory because the paramagnetic SSE was observed at low $T$ and large $B$. We replaced $B$ with the Curie-Weiss molecular (effective) magnetic field \cite{hu2014effective, daudin1982thermodynamic} $B_{\mathrm{eff}} = \left[ T/(T - \varTheta_{\mathrm{CW}})\right]B$, where $\varTheta_{\mathrm{CW}}$ is the Curie-Weiss temperature of the considered paramagnet to calculate Eqs. (\ref{Bf}) and (\ref{Bf2}), when we apply our model to a specific case such as the Pt/GGG interface in Sec. \ref{compari}. {In the case of ferromagnetic exchange interaction ($\Theta_{\rm CW} > 0$), our approximation is only valid in $T> \varTheta_{\mathrm{CW}}$, because $B_{\mathrm{eff}}$ diverges at $T = \varTheta_{\mathrm{CW}}$. By contrast, in the case of antiferromagnetic exchange interaction, $\varTheta_{\mathrm{CW}}$ is negative, and $B_{\mathrm{eff}}$ is always well defined and gives quantitative calculation for $\xi \alt 1$. }

\subsection{Spin current at NM/PI interface}
We formulate the spin-current density $j_{\mathrm{s}}$ at the interface with the interfacial exchange interaction between the conduction electron in NM and localized spin in PI based on the linear response theory for the spin current in the NM/FM junction \cite{takahashi2010spin}. The interfacial exchange interaction Hamiltonian reads 

\begin{equation}
\mathcal{H}_{\mathrm{int}} = -\nu_{\mathrm{N}} \mathcal{J}_{\mathrm{int}} \Sigma^{N_{\mathrm{PI}}}_{n=1} \bold{S}_{{n}} (t) \cdot \boldsymbol{\sigma}(\bold{r}_n, t),
\end{equation}
where $\mathcal{J}_\mathrm{int}$ is the coefficient of the interfacial exchange interaction, $\bold{S}_n(t)$ is the local spin at the position $\bold{r}_n$ and time $t$, $\boldsymbol{\sigma}(\bold{r}_n, t) = 2\bold{s}(\bold{r}_n, t) $ is the conduction electron spin density at $\bold{r}_n$, $\nu_{\mathrm{N}}$ is the unit cell volume of NM, and $N_{\mathrm{PI}}$ is the number of local spins at the interface. 

The spin-current density operator polarized in the $z$ direction and flowing in the $y$ direction at the interface $\hat{j}^{z}_{\mathrm{s}} (t) = (\hbar/2A) d( \hat{N}^{\uparrow}_e - \hat{N}^{\downarrow}_e )/dt $, where $\hat{N}^{\alpha}_e$ is the number operator of the conduction electron with spin $\alpha$ ($\alpha = \uparrow , \downarrow$) and $A$ is the area size of the interface, is calculated by the Heisenberg equation of motion as 
\begin{equation}
\hat{j}^{z}_{\mathrm{s}} (t) = - \frac{\nu_{\mathrm{N}} \mathcal{J}_{\mathrm{sf}} }{A} \Sigma^{N_{\mathrm{PI}}}_{n =1} \left[ \bold{S}_n (t) \times \sigma(r_n) \right]_z.
\end{equation}
We calculate the second-order perturbation of NM by the interfacial exchange interaction using the linear response theory and obtain the spin current density $j_{\mathrm{s}}$ across the interface as 
\begin{align}
\label{spincurrent}
j_{\mathrm{s}} &= -\frac{i}{\hbar} \int^{t}_{-\infty} dt' \left< \left[ \hat{j}^{z}_{\mathrm{s}} (t'), \hat{\mathcal{H}}_{\mathrm{sf}}(t') \right] \right> \notag \\
&= \frac{n_{\mathrm{PI}}}{\hbar} (\nu_{\mathrm{N}} \mathcal{J}_{\mathrm{sf}})^2 \int^{\infty}_{-\infty} dt \Sigma_{p} \left[ \langle S_{+}(t) S_{-}(0) \rangle \langle \sigma_{-p}^{-} (t) \sigma_{p}^{+}(0) \rangle \right. \notag  \\
& \left. \quad \quad \quad - \langle S_{-}(t) S_{+}(0)\rangle \langle \sigma_{p}^{+}(t) \sigma_{-p}^{-}(0) \rangle \right],
\end{align}
where spin $S$ is a representative of $S_n$ at the interface{, $n_{\mathrm{PI}}$ is the interfacial spin density,} and $S_{\pm} = S_{x} \pm i S_{y}$ is the ladder operator for the local spin with the quantized axis chosen along $z$, $\sigma^{+}_{p} = (1/V_{\mathrm{N}}) \Sigma_{k} c^{\dagger}_{k\uparrow} c_{k+p\downarrow}$ and $\sigma^{-}_{-p} = (1/V_{\mathrm{N}}) \Sigma_{k} c^{\dagger}_{k+p\uparrow} c_{k\downarrow}$ are the transverse spin density of the conduction electrons in NM, and $c_{k \sigma}(t) = c_{k\sigma} \mathrm{exp}{(-i\epsilon_{k} t)}$ and $c^{\dagger}_{k \sigma}(t) = c^{\dagger}_{k\sigma} \mathrm{exp}{(i\epsilon_{k} t)}$ are the creation and annihilation operator of an electron with momentum $k$ and spin $\sigma$, respectively, and $\epsilon_k$ is the one-electron energy.

The spin-spin correlation function of the localized spin in PI can be written as
\begin{equation}
\label{Spm}
\langle S_{+}(t) S_{-}(0) \rangle = \langle (S+m)(S-m+1) \rangle \mathrm{exp}({+ig\mu_{\mathrm{B}} B/\hbar}), 
\end{equation}
\begin{equation}
\label{Smp}
\langle S_{-}(t) S_{+}(0)\rangle  = \langle (S-m)(S+m+1) \rangle \mathrm{exp}({-ig\mu_{\mathrm{B}}B /\hbar}),
\end{equation}
where $m$ is a quantum number of $S_{z}$, which we call the magnetization along $\bold{B}$ of the paramagnetic spins in PI.
We take $\mu_{\mathrm{PI}}$ into account for calculating $\langle m \rangle$ and $\langle m^{2} \rangle$ using Eqs. (\ref{Bf}) and (\ref{Bf2}) to consider the spin-current-induced nonequilibrium spin states in PI.
By putting Eqs. (\ref{Spm}) and (\ref{Smp}) into Eq. (\ref{spincurrent}), $j_{\mathrm{s}}$ becomes
\begin{align}
j_{\mathrm{s}}&= \frac{n_{\mathrm{PI}}}{\hbar} J_{\mathrm{int}}^2 \langle (S+m)(S-m-1) \rangle \Pi_{-+} \notag \\
&\quad \quad - \langle (S- m)(S+m+1) \rangle \Pi_{+-},
\end{align}  
where $\Pi_{-+} = \int^{\infty}_{-\infty} dt \Sigma_{p} \langle \sigma_{-p}^{-}(t) \sigma_{p}^{+}(0) \rangle \mathrm{e}^{+ig\mu_{\mathrm{B}}B/\hbar}$ and $\Pi_{+-} = \int^{\infty}_{-\infty} dt \Sigma_{p} \langle \sigma_{p}^{+}(t) \sigma_{-p}^{-}(0) \rangle \mathrm{e}^{-ig\mu_{\mathrm{B}}B/\hbar} $.
The spin-spin correlation function of the conduction-electron part can be calculated as
\begin{align}
\Pi_{-+} &= 2\pi \hbar N^2(0) (\hbar \omega_B - \mu_{\mathrm{NM}})\left[ n_{\mathrm{B}}(\hbar \omega_B - \mu_{\mathrm{NM}}, T_{\mathrm{NM}}) +1 \right], \\
\Pi_{+-} &= 2\pi \hbar N^2(0) (\hbar \omega_B - \mu_{\mathrm{NM}})\left[ n_{\mathrm{B}}(\hbar \omega_B - \mu_{\mathrm{NM}}, T_{\mathrm{NM}}) \right],
\end{align}
where $N(0)$ is the density of state at the Fermi level and $\omega_B = g\mu_{\mathrm{B}}B/\hbar$ is the Lamor frequency, and $n_\mathrm{B} (\epsilon,T) $ is the Bose distribution function with energy $\epsilon$ and temperature $T$.
Finally, we obtain 
\begin{align}
\label{spincurrentfinal}
j_{\mathrm{s}} &= {2}j_{\mathrm{s}}^0 \left(\frac{\hbar \omega_B - \mu_{\mathrm{NM}}}{k_{\mathrm{B}} T_{\mathrm{NM}}} \right) SB_S \left( \frac{\hbar \omega_B - \mu_{\mathrm{PI}}}{k_{\mathrm{B}} T_{\mathrm{PI}}} \right) \notag \\
& \times  \left[ n_{\mathrm{B}}(\hbar \omega_B - \mu_{\mathrm{PI}}, T_{\mathrm{PI}}) - n_{\mathrm{B}}(\hbar \omega_B - \mu_{\mathrm{NM}}, T_{\mathrm{NM}}) \right], 
\end{align}
where $j_{\mathrm{s}}^0 = {2}\pi n_{\mathrm{PI}} J_{\mathrm{int}}^2 k_{\mathrm{B}}T_{\mathrm{NM}}$ and $\nu_{\mathrm{N}} \mathcal{J}_{\mathrm{sf}} N(0) = J_{\mathrm{int}}$ is the dimensionless interfacial exchange interaction. 
We found that $j_{\mathrm{s}}$ is proportional to the thermal average of the magnetization of PI ($SB_{S}$) and the difference between the distribution function of PI and NM. 
Because the spin state is labelled by the effective temperature $T_{\mathrm{X}}$ and spin chemical potential $\mu_{\mathrm{X}}$ (X = NM or PI), Eq. (\ref{spincurrentfinal}) indicates the finite spin current appears when $T_{\mathrm{X}}$ and/or $\mu_{\mathrm{X}}$ between NM and PI are different. {The difference of $T_{\mathrm{X}}$ and $\mu_{\mathrm{X}}$ between NM and PI arises, for example, when the temperature gradient is applied to the system and SHE creates spin accumulation in the NM side by the application of the current.} It is worth mentioning that Eq. (\ref{spincurrentfinal}) gives the general form of the interfacial spin current at NM/PI, which can be applied to the SSEs as well as the nonlocal spin transport \cite{oyanagi2019spin} and spin Hall magnetoresistance \cite{zhang2019theory, oyanagi2021paramagnetic} with paramagnetic insulators.

\subsection{Application to SSE in NM/PI system}
In this subsection, we apply Eq. (\ref{spincurrentfinal}) to describe the SSE in the NM/PI system by calculating the interfacial spin current in Eq. (\ref{spincurrentfinal}) under a temperature difference between PI and NM ($T_{\mathrm{NM}}/T_{\mathrm{PI}} \neq 1$). By application of {a small temperature difference $\Delta T = T_{\mathrm{NM}} - T_{\mathrm{PI}} \ll T_{\mathrm{PI}}$}, an imbalance of the thermal spin pumping from PI and the back-flow spin current from NM appears, resulting in a finite flow of spin current at the interface. In NM, the resultant spin current creates the spin accumulation, which diffuses and is converted into a voltage via the ISHE. 

Figure \ref{fig2c}(b) shows the calculation of $j_{\mathrm{s}}/j^{0}_{\mathrm{s}}$ as a function of $\xi = g \mu_{\mathrm{B}}B/k_{\mathrm{B}}T_{\mathrm{PI}}$ at various $S$ for $T_{\mathrm{NM}}/T_{\mathrm{PI}} = 0.8$. To focus on the effect of the temperature difference, we set $\mu_{\mathrm{NM,PI}} = 0$ in the calculation. The calculated spin current changes the sign by reversing the sign of $\xi$. 
The magnitude of $j_{\mathrm{s}}/j^{0}_{\mathrm{s}}$ increases for $\left|\xi \right| < 1$ and takes a maximum value at $\left|\xi \right| \thickapprox 1$, while it decreases for $\left|\xi \right| > 1$.
$j_{\mathrm{s}}/j^{0}_{\mathrm{s}}$ disappears at $\left|\xi \right| = 0 $.
With the increase in $S$, the maximum value of $j_{\mathrm{s}}/j^{0}_{\mathrm{s}}$ monotonically increases. 
The $\xi$ dependence of $j_{\mathrm{s}}/j^{0}_{\mathrm{s}}$ in $\left|\xi \right| \alt 1$ is consistent with that of $\langle m\rangle$ [see Eq. (\ref{spincurrentfinal}) and Fig. \ref{fig2c}(a)], indicating the aligning of the localized spins in PI by the $B$-induced increase in $j_{\mathrm{s}}/j^{0}_{\mathrm{s}}$.
The decrease of $j_{\mathrm{s}}/j^{0}_{\mathrm{s}}$ is substantial when the Zeeman energy exceeds thermal energy ($\left|\xi \right| > 1$) because of the suppression of the spin-flip scattering at the interface [see Fig. \ref{fig1}(b)].
Our result indicates that the finite magnetization is responsible for the generation of the interfacial spin current, and the overall shape of $j_{\mathrm{s}}/j^{0}_{\mathrm{s}}$ is related to the $B$-induced alignment of localized spin and Zeeman gap opening.

We next describe the effect of a finite spin accumulation $\mu_{\mathrm{NM}}$ on the SSE. The thermally generated spin current flows across the interface and creates $\mu_{\mathrm{NM}}(y)$ in NM, which diffuses to flow in NM, $j^{\mathrm{NM}}_{\mathrm{s}}(y)$.
Due to the ISHE in NM, $j^{\mathrm{NM}}_{\mathrm{s}}(y)$ generates an electric field $\bold{E}_{\mathrm{ISHE}}$ [see Fig. \ref{fig2-2}] given by $E_{\mathrm{ISHE}} = \theta_{\mathrm{SHE}}\rho_{\mathrm{NM}} (2e/\hbar) \langle j^{\mathrm{NM}}_{\mathrm{s}}(y) \rangle$, where $\theta_{\mathrm{SHE}}$ is the spin Hall angle of NM, $\rho_{\mathrm{NM}}$ is the resistivity of NM, $e$ is the elementary charge, and $\langle j^{\mathrm{NM}}_{\mathrm{s}}(y) \rangle$ is the average of $j^{\mathrm{NM}}_{\mathrm{s}}(y)$ over $y$. By solving the spin diffusion equation for $\mu_{\mathrm{NM}}(y) $ in NM with the spin-current continuity boundary condition, the ISHE-induced voltage, $V_{\mathrm{SSE}}$, becomes 
\begin{figure*}
\includegraphics{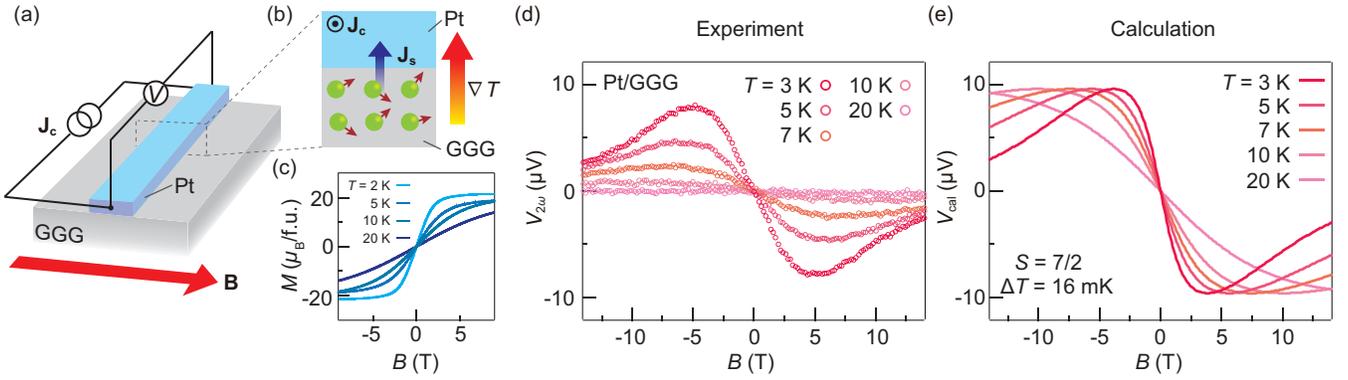}
\caption{\label{fig3e} A schematic illustration of the measurement setup of the paramagnetic spin Seebeck effect in the Pt/GGG system (a) and the magnified view of the Pt/GGG interface (b). $\bold{J}_\mathrm{c}$, $\bold{B}$, $\bold{J}_{\mathrm{s}}$, and $\nabla T$ show the applied charge current, external magnetic field, interface spin current, and Joule-heating induced temperature gradient, respectively. (c) $M(B)$ of GGG at various $T$. (d) Experimental results of the $B$ dependence of $V_{2\omega}$ in the Pt/GGG system at various $T$. (d) Theoretical calculation of the $B$ dependence of the paramagnetic spin Seebeck voltage $V_{\mathrm{cal}}$ at various $T$ with $\Delta T = 16$ mK and the material parameters summarized in Table I.}
\end{figure*}
\begin{align}
\label{voltage}
V_{\mathrm{SSE}} = \frac{2e}{\hbar} \theta_{\mathrm{SH}} \rho_{\mathrm{NM}} \lambda_{\mathrm{NM}}\frac{l_{\mathrm{NM}}}{d_{\mathrm{NM}}} \mathrm{tanh}\left( \frac{d_{\mathrm{NM}}}{2\lambda_{\mathrm{NM}}}\right) j_{\mathrm{s}},
\end{align}
where $\lambda_{\mathrm{NM}}$ is the spin diffusion length of NM, $l_{\mathrm{NM}}$ is the length of the NM contact, and $d_{\mathrm{NM}}$ is the thickness of NM.
Expanding $j_{\mathrm{s}}$ in Eq. (\ref{spincurrentfinal}) up to the linear order in both $\mu_{\mathrm{NM}}(0)$ and $\Delta T$, and using the solution for the spin diffusion equation, we obtain the interfacial spin current as
\begin{equation}
\label{interfacespincurrent}
j_{\mathrm{s}} = -S_{\mathrm{SSE}} k_{\mathrm{B}}\Delta T,
\end{equation}
where $S_{\mathrm{SSE}}$ is the spin Seebeck conefficient
\begin{equation}
S_{\mathrm{SSE}} = \frac{\hbar}{2e^{2}} \frac{g \mu_{\mathrm{B}}B}{2k_{\mathrm{B}} T}
\frac{2g_{\mathrm{s}}}{1+2\rho_{\mathrm{NM}} \lambda_{\mathrm{NM}}g_{\mathrm{s}}\mathrm{coth}(d_{\mathrm{NM}}/\lambda_{\mathrm{NM}})}, 
\end{equation}
with the effective spin conductance \cite{oyanagi2019spin, zhang2019theory, oyanagi2021paramagnetic},
\begin{align}
g_{\mathrm{s}} = 2\pi \frac{2e^{2}}{\hbar} n_{\mathrm{PI}}J^{2}_{\mathrm{int}} SB_{S}\left( \frac{g \mu_{\mathrm{B}}B}{2k_{\mathrm{B}} T} \right) \left[ \frac{(g \mu_{\mathrm{B}}B/2k_{\mathrm{B}} T )}{\mathrm{sinh}^{2}(g \mu_{\mathrm{B}}B/2k_{\mathrm{B}} T )} \right],
\end{align}
for small temperature difference ($T_{\mathrm{NM}}/T_{\mathrm{PI}} \sim 1$) and small spin conductance condition ($\rho_{\mathrm{NM}}\lambda_{\mathrm{NM}}g_{\mathrm{s}} \ll 1$), which is always satisfied at sufficiently low temperatures.

\section{EXPERIMENTAL RESULTS AND COMPARISON WITH THEORY}
\label{compari}
In this section, we show that our model well explains the observed paramagnetic SSE signal in the Pt/GGG sample [Figs. \ref{fig3e}(a) and \ref{fig3e}(b)]. We observed the paramagnetic SSE in the low temperatures ($T < 20$ K) and high magnetic field ($B > 1$ T) regime similar to the previous reports \cite{wu2015paramagnetic, liu2018probing}.

\subsection{Magnetization of GGG}
Paramagnetic insulator GGG is commonly used as a substrate for growing a thin-film ferrimagnetic Y$_3$Fe$_5$O$_{12}$ (YIG) \cite{sun2012growth} in spintronics and magnonics \cite{serga2010yig, chumak2015magnon}. Since the Ga$^{3+}$ ion is nonmagnetic, the magnetic property of GGG is governed by the Gd$^{3+}$ ions with the small antiferromagnetic exchange interaction of $-0.1$ K \cite{schiffer1995frustration}. GGG shows no long-range magnetic ordering down to 180 mK and has a small Curie-Weiss temperature $\varTheta_{\mathrm{CW}}$ of $-2$ K \cite{schiffer1995frustration}, making GGG an ideal classical paramagnetic system.

Figure \ref{fig3e}(c) shows the $B$ dependence of GGG's magnetization, $M$, measured by a vibrating sample magnetometer at various $T$. The saturation magnetization of $\sim$21 $\mu_{\mathrm{B}}/\mathrm{f.u.}$ is consistent with the expected value for Gd$_3$Ga$_5$O$_{12}$, where Gd$^{3+}$ ($S=7/2$) carries $\sim$ 7$\mu_{\mathrm{B}}$. The $M-B$ curve at the lowest $T$ of 3 K shows a Brillouin-function-like response; $M$ is saturated at large $ B$. \ref{fig2c}(a). By contrast, $M$ at 20 K increases linearly with $B$ and resembles the calculated $M$ for $ \xi \ll 1$.
$M-B$ at low (high) $T$ corresponds to the curves in the large (small) $\xi$ region in Fig. \ref{fig2c}(a). 
All magnetization measurement results indicate GGG obeys the Curie-Weiss law down to low temperature.

\subsection{Experimental setup and results}
We fabricated an on-chip spin Seebeck device \cite{wu2015spin, wu2015paramagnetic, wu2016antiferromagnetic, kikkawa2021observation} schematically shown in Fig. \ref{fig3e}(a).
The device consists of a Pt strip on a GGG slab ($10\times10\times0.5$ mm$^3$) commercially obtained from CRYSTAL GmbH. Before the nanofabrication, we cleaned the GGG slab in acetone for 5 minutes using an ultrasonication bath. We picked up it from acetone and blew off the acetone residue on top of the GGG slab with dried nitrogen gas. We used positive resist PMMA 950A and conductive organic polymer ESPACER (Showa Denko) for preventing charge up during the $e$-beam lithography. The dimension of Pt is 200 $\mu$m long, 100 nm wide, and 10 nm thickness prepared by magnetron sputtering, $e$-beam lithography and lift-off methods \cite{cornelissen2017nonlocal, oyanagi2019spin, oyanagi2020magnetic}. We confirmed that the surfaces of the GGG substrate and Pt film have small surface roughnesses using an atomic force microscope [please see Figs. 6(a) and (b) in Appendix A]. The Pt strip works as a heater as well as a spin-current detector. 
We generate Joule heating to induce a thermal gradient $\nabla T$ across the Pt/GGG interface by applying a current $\bold{J}_\mathrm{c}$ with magnitude $J_{\mathrm{c}}$. The generated $\nabla T$ drives a spin current $\bold{J}_{\mathrm{s}}$ at the interface via the exchange interaction between localized spins in GGG and conduction electron spins in Pt [see Fig. \ref{fig3e}(b)].
The spin current is converted into a voltage signal via the ISHE in Pt. We applied $J_{\mathrm{c}}$ with the frequency of 13.423 Hz and root-mean-square amplitude of 10 $\mu$A, which corresponds to the heating power of $\sim 10$ $\mu$W. We fixed the heating power instead of the temperature difference in the SSE measurements, which is different from the theoretical calculation with the input of the constant temperature difference. We confirmed that the system temperature remains unchanged during the SSE measurements, indicating the temperature increase due to Joule heating is negligible.
As the thermally induced spin current has the same frequency of the Joule heating ($\propto J^2_{\mathrm{c}}$), we measured the second harmonic voltage $V_{2\omega}$ across the Pt strip using a lock-in method to resolve the resultant SSE signals.

Figure \ref{fig3e}(d) shows the $B$ dependence of $V_{2\omega}$ in the Pt/GGG sample for various $T$. At $T=20$ K, we observed no voltage signal. With decreasing $T$, a clear $V_{2\omega}$ appears, whose sign changes with respect to the $B$ direction. 
This is due to the reversal of the direction of the localized spins in GGG and the spin-current polarization. 
The maximum signal intensity rapidly increases with decreasing $T$. $V_{2\omega}(B)$ below 10 K linearly increases up to $B=2$ T and takes the maximum value at around $B=5$ T. 
By further increasing $B$, $V_{2\omega}$ starts to decrease, showing the $B$-induced reduction of the paramagnetic SSE. All observation here is consistent with the results of the paramagnetic SSE reported by Wu $et$ $al.$ \cite{wu2015paramagnetic}. 

\subsection{Calculation of spin Seebeck voltage in Pt/GGG}
{We numerically calculated the paramagnetic spin Seebeck voltage $V_{\mathrm{cal}}$ in the Pt/GGG system given by Eq. (\ref{voltage}) with the material parameters summarized in Table I using Microsoft Excel}. Figure \ref{fig3e}(e) shows $V_{\mathrm{cal}}(B)$ under the constant temperature difference of $\Delta T = T_{\mathrm{Pt}} - T_{\mathrm{GGG}} =16$ mK at the selected $T$. At all $T$, $V_{\mathrm{cal}}$ appears with the application of $B$ and changes the sign under the $B$ reversal. At $T \le$ 10 K, $V_{\mathrm{cal}}$ increases with $B < 5$ T and shows a broad peak at around 5 T. In the $B > 5$ T range, we found the $B$-induced reduction of $V_{\mathrm{cal}}$. By contrast, $V_{\mathrm{cal}}$ at $T >$ 10 K monotonically increases up to $B=14$ T. Note that the $B$ dependence at low (high) $T$ corresponds to the $\xi$ dependence of $j_{\mathrm{s}}/j^{0}_{\mathrm{s}}$ for $\xi > 1$ ($\xi < 1$) [see Fig. \ref{fig2c}(b)].

\begin{figure}[t]
\includegraphics{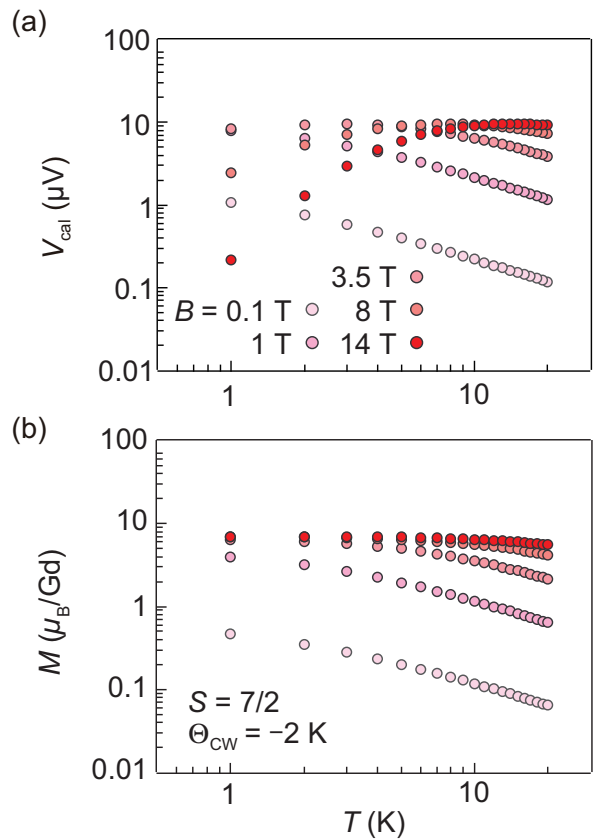}
\caption{\label{fig4b} Calculation of the paramagnetic spin Seebeck voltage $V_{\mathrm{cal}}(T)$ for Pt/GGG (a) and {bulk magnetization $M(T)$ of GGG (b) with the bulk Curie-Weiss temperature of -2 K} at selected $B$. }
\end{figure}

Next, we investigated the $T$ dependence of $V_{\mathrm{cal}}$. Figure \ref{fig4b}(a) shows $V_{\mathrm{cal}}(T)$ at various $B$ values.
{Except for $V_{\mathrm{cal}}(T)$ at $B=14$ T, $V_{\mathrm{cal}}(T)$ monotonically increases in $5$ K $\le T \le 20$ K. This behavior is similar to $M(T)$, where $M$ follows the Curie-Weiss law [$M \propto (T-\Theta_{\mathrm{CW}})^{-1}$] shown in Fig. \ref{fig4b}(b)}. Below 5 K, $V_{\mathrm{cal}}$ at {$B < 3.5$ T} increases down to 1 K, while {$V_{\mathrm{cal}}(T)$} at {$B \geq 3.5$ T} decreases, corresponding to the $B$-induced reduction of $V_{\mathrm{cal}}$. {This is different from the saturation behavior of $M$ in the low $T$ and high $B$ range. }

\subsection{Comparison between theory and experiment}
By comparing Figs. \ref{fig3e}(d) and \ref{fig3e}(e), we found that the calculation well reproduces the experimental $B$ dependence of the paramagnetic SSE. The agreement between the calculation and experiments indicates that the $B$-induced reduction of the paramagnetic SSE is ascribed to the competition between the Zeeman energy ($\propto g\mu_{\mathrm{B}}B$) and thermal energy ($\propto k_{\mathrm{B}}T$). When sufficiently large $B$, such as $k_{\mathrm{B}}T \ll g\mu_{\mathrm{B}}B$, is applied, the spin-flip scattering reduces because the thermal energy cannot overcome the Zeeman gap [see Fig. \ref{fig1}(b)], and thus the interfacial spin current and SSE signal reduce. The same mechanism is also responsible for the $B$-induced reduction of the SSE in FMs at low $T$, where freeze-out of magnons prohibits the thermal magnon excitation \cite{kikkawa2015critical, jin2015effect, kikkawa2016complete, oyanagi2020magnetic}. 

\begin{table}[btp]
  \caption{Selected parameters for calculating the paramagnetic spin Seebeck voltage $V_{\mathrm{cal}}$ in Pt/GGG. $\Theta^{\mathrm{int}}_{\mathrm{CW}}$, $J_{\mathrm{int}}$, and $n_{\mathrm{PI}}$ are estimated in Ref.\cite{oyanagi2021paramagnetic} using $\lambda_{\mathrm{Pt}}$, $\Theta_{\mathrm{SH}}$, and $\rho_{\mathrm{Pt}}$ in Ref. \cite{sagasta2016tuning}.} 
  \label{table:parameter}
  \begin{tabular*}{8.65cm}{@{\extracolsep{\fill}}lccc}
    \hline\hline
       & Symbol  &  Value &  Unit \\
    \hline 
    Platinum spin diffusion length&$\lambda_{\mathrm{Pt}}$&2&nm\\
    Platinum spin Hall angle&$\theta_{\mathrm{SH}}$&0.11 &-\\
    Platinum resistivity&$\rho_{\mathrm{Pt}}$&3.4$\times10^{-7}$&$\Omega$m\\
    Gd spin angular momentum&$S$&7/2&-\\
    \begin{tabular}{c}
    Interfacial Curie-Weiss\\temperature\end{tabular}&$\Theta^{\mathrm{int}}_{\mathrm{CW}}$& -1.27&K\\
    \begin{tabular}{c}
    Dimensionless interfacial \\exchange interaction  
     \end{tabular} &$J_{\mathrm{int}}$&-0.065& - \\
    Interfacial Gd atom density&$n_{\mathrm{PI}}$& 6.94$\times10^{16}$  &Gd/m$^{2}$\\
    \hline\hline
  \end{tabular*}
\end{table}
{We found that $V_{\mathrm{cal}}$ and $V_{\mathrm{2\omega}}$ show good agreement in the amplitude at 3 K with the input of $\Delta T = 16$ mK and the material parameters in Table I. This value is comparable to the increase of the Pt surface temperature estimated in the similar Pt/GGG device at 2 K using the thermal simulation \cite{liu2018probing}. The interfacial temperature difference originates from an effective Kapitza conductance $\kappa_{\mathrm{eff}}$, which relates the injected heat flux to $\Delta T$ at the interface \cite{reitz2020spin}. We obtained $\kappa_{\mathrm{eff}} \sim 1.5\times 10^{7}$ Wm$^{-2}$K$^{-1}$ at 3 K, which reasonably agrees to that in Pt/YIG at low temperatures of 40 K \cite{angeles2021interfacial}. However, it may also indicate the possible overestimation of $\kappa_{\mathrm{eff}}$ because of the strong decrease of $\kappa_{\mathrm{eff}}$ at lower temperatures \cite{kent1993experimental}. For more quantitative comparison, we need detailed experiments on the interfacial heat and spin transport at low temperatures.}

Finally, we discuss the $T$ dependence of the paramagnetic SSE. {The interface spin current takes the maximum at $\xi \thickapprox 1$ shown in Fig. 3(b), and so does $V_{\mathrm{cal}}$ at $T_{\mathrm{max}} \thickapprox g\mu_{\mathrm{B}}B_{\mathrm{eff}}/k_{\mathrm{B}} = (g\mu_{\mathrm{B}}B/k_{\mathrm{B}}) - |\Theta_{\mathrm{CW}}|$. Quantitatively, $V_{\mathrm{cal}}$ at $B=3.5$ T shows the $T^{-0.6}$ dependence in $5$ K $\alt T \alt 20$ K, above $T_{\mathrm{max}}(B = 3.5\ \mathrm{T}) \thickapprox 3.4$ K}. However, the experimental results reported by Wu $et$ $al.$ \cite{wu2015paramagnetic} show the faster power-law decay of the signal, $T^{-3.4}$, in the Pt/GGG system in the condition. We also obtained a similar $T$ dependence in our experimental results. The difference can be caused by the $T$ dependence of $\Delta T$. In our calculation, we fixed the constant $\Delta T$ value of {16 mK} for obtaining $V_{\mathrm{cal}}$, corresponding to the intrinsic $T$ dependence of the paramagnetic SSE. However, in the experiments, actual $\Delta T$ is not constant with varying $T$, even though the heating power is fixed. At low $T$, the thermal conductivity of GGG strongly decreases with decreasing $T$ \cite{daudin1982thermodynamic}, causing the $T$ dependence of $\Delta T$, which affects the $T$ dependence of the paramagnetic SSE voltage signal. Indeed, Wu $et$ $al.$ obtained the $T$ dependence of the paramagnetic SSE voltage signal $\propto T^{-4}$ by taking the $T$ dependence of the thermal conductivity of GGG, the Kapitza conductance at the interface, and $M$ of GGG into account. Comparison with our calculation and experimental results measured with constant $\Delta T$ \cite{iguchi2017concomitant} may provide further insight into the $T$ dependence of paramagnetic SSE in this system.

\section{SUMMARY}
We theoretically investigated the spin Seebeck effect (SSE) in a normal metal (NM)/paramagnetic insulator (PI) junction. The spin and energy exchange appears at the NM/PI interfacial due to the spin-flip scattering between the conduction electron spins in NM and localized spins in PI. We calculated the spin current density at the interface using the linear response theory and found that the finite spin current appears when the temperatures of spins in NM and PI are different. 
The interfacial spin-current density is governed by values of the electron spin $S$ of the localized spin in PI and $\xi \propto B/T$.
When $\xi < 1$, the spin current density increases due to the alignment of the localized spin, while the spin current density is rapidly suppressed when $\xi > 1$ because the Zeeman energy exceeds thermal energy, resulting in the suppression of the spin-flip scattering. 
The good agreement between the calculated and measured $B$ dependences of the paramagnetic SSE voltage in the Pt/GGG system at low temperatures indicates that the $B$-induced reduction of the paramagnetic SSE is attributed to the suppression of the interfacial spin-flip scattering due to the Zeeman gap opening. {Recent studies on the long-rage spin transport in paramagnets \cite{oyanagi2019spin,luo2022spin} indicate the importance of bulk spin transport, and thus future work on a SSE theory based on bulk spin and heat transport in paramagnets is important to clarify the role of spin-wave excitation mediated by the dipole interaction for the paramagnetic SSE.} Our results clarify the mechanism of the thermally induced spin transport at NM/PI interfaces and give insight into the thermal spin current generation in spin caloritronics \cite{bauer2012spin}. 

\section*{ACKNOWLEDGMENT}
We thank S. Daimon, G. E. W. Bauer and B. J. van Wees for fruitful discussion. This work is a part of the research program of ERATO “Spin Quantum Rectification Project” (No. JPMJER1402) and CREST (Nos. JPMJCR20C1 and JPMJCR20T2) from JST, the Grant-in-Aid for Scientific Research on Innovative Area “Nano Spin Conversion Science” (No. JP26103005), the Grant-in-Aid for Scientific Research (S) (No. JP19H05600), Grant-in-Aid for Scientific Research (B) (No. JP20H02599), Grant-in-Aid for Research Activity start-up (No. JP20K22476), Grant-in-Aid for Early-Career Scientists (No. 21K14519) from JSPS KAKENHI, JSPS Core-to-Core program, the International Research Center for New-Concept Spintronics Devices, World Premier International Research Center Initiative (WPI) from MEXT, Japan, and Institute for AI and Beyond of the University of Tokyo. K.O. acknowledges support from GP-Spin at Tohoku University.

\section*{Appendix A: Surface of GGG and Pt film on GGG}
Figures 6(a) and 6(b) show the atomic force microscope (AFM) images of the surface of the GGG substrate after the cleaning and Pt film on GGG. We found similar values of the surface roughness: $R_{\mathrm{a}}$ = 0.3 nm for the GGG substrate and $R_{\mathrm{a}}$ = 0.4 nm for the Pt film. $R_{\mathrm{a}}$ of 0.4 nm is much smaller than the Pt thickness, indicating high quality and uniformity of the film.

\begin{figure}[h]
\includegraphics{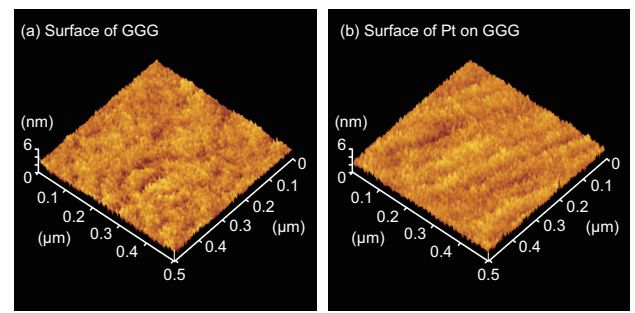}
\caption{\label{fig6b} {Atomic force microscope images of the polished (111) surface of GGG (a) and the Pt film on the GGG substrate (b).}}
\end{figure}

\bibliography{citation2}

\begin{thebibliography}{63}
\expandafter\ifx\csname natexlab\endcsname\relax\def\natexlab#1{#1}\fi
\expandafter\ifx\csname bibnamefont\endcsname\relax
  \def\bibnamefont#1{#1}\fi
\expandafter\ifx\csname bibfnamefont\endcsname\relax
  \def\bibfnamefont#1{#1}\fi
\expandafter\ifx\csname citenamefont\endcsname\relax
  \def\citenamefont#1{#1}\fi
\expandafter\ifx\csname url\endcsname\relax
  \def\url#1{\texttt{#1}}\fi
\expandafter\ifx\csname urlprefix\endcsname\relax\def\urlprefix{URL }\fi
\providecommand{\bibinfo}[2]{#2}
\providecommand{\eprint}[2][]{\url{#2}}

\bibitem[{\citenamefont{Uchida et~al.}(2008)\citenamefont{Uchida, Takahashi,
  Harii, Ieda, Koshibae, Ando, Maekawa, and Saitoh}}]{uchida2008observation}
\bibinfo{author}{\bibfnamefont{K.}~\bibnamefont{Uchida}},
  \bibinfo{author}{\bibfnamefont{S.}~\bibnamefont{Takahashi}},
  \bibinfo{author}{\bibfnamefont{K.}~\bibnamefont{Harii}},
  \bibinfo{author}{\bibfnamefont{J.}~\bibnamefont{Ieda}},
  \bibinfo{author}{\bibfnamefont{W.}~\bibnamefont{Koshibae}},
  \bibinfo{author}{\bibfnamefont{K.}~\bibnamefont{Ando}},
  \bibinfo{author}{\bibfnamefont{S.}~\bibnamefont{Maekawa}}, \bibnamefont{and}
  \bibinfo{author}{\bibfnamefont{E.}~\bibnamefont{Saitoh}},
  \bibinfo{journal}{Nature (London)} \textbf{\bibinfo{volume}{455}},
  \bibinfo{pages}{778} (\bibinfo{year}{2008}).

\bibitem[{\citenamefont{Uchida et~al.}(2014{\natexlab{a}})\citenamefont{Uchida,
  Ishida, Kikkawa, Kirihara, Murakami, and
  Saitoh}}]{uchida2014longitudinalreview}
\bibinfo{author}{\bibfnamefont{K.}~\bibnamefont{Uchida}},
  \bibinfo{author}{\bibfnamefont{M.}~\bibnamefont{Ishida}},
  \bibinfo{author}{\bibfnamefont{T.}~\bibnamefont{Kikkawa}},
  \bibinfo{author}{\bibfnamefont{A.}~\bibnamefont{Kirihara}},
  \bibinfo{author}{\bibfnamefont{T.}~\bibnamefont{Murakami}}, \bibnamefont{and}
  \bibinfo{author}{\bibfnamefont{E.}~\bibnamefont{Saitoh}},
  \bibinfo{journal}{J. Phys. Condens. Matter} \textbf{\bibinfo{volume}{26}},
  \bibinfo{pages}{343202} (\bibinfo{year}{2014}{\natexlab{a}}).

\bibitem[{\citenamefont{Uchida et~al.}(2016)\citenamefont{Uchida, Adachi,
  Kikkawa, Kirihara, Ishida, Yorozu, Maekawa, and
  Saitoh}}]{uchida2016thermoelectric}
\bibinfo{author}{\bibfnamefont{K.-i.} \bibnamefont{Uchida}},
  \bibinfo{author}{\bibfnamefont{H.}~\bibnamefont{Adachi}},
  \bibinfo{author}{\bibfnamefont{T.}~\bibnamefont{Kikkawa}},
  \bibinfo{author}{\bibfnamefont{A.}~\bibnamefont{Kirihara}},
  \bibinfo{author}{\bibfnamefont{M.}~\bibnamefont{Ishida}},
  \bibinfo{author}{\bibfnamefont{S.}~\bibnamefont{Yorozu}},
  \bibinfo{author}{\bibfnamefont{S.}~\bibnamefont{Maekawa}}, \bibnamefont{and}
  \bibinfo{author}{\bibfnamefont{E.}~\bibnamefont{Saitoh}},
  \bibinfo{journal}{Proc. IEEE} \textbf{\bibinfo{volume}{104}},
  \bibinfo{pages}{1946} (\bibinfo{year}{2016}).

\bibitem[{\citenamefont{Uchida}(2021)}]{uchida2021transport}
\bibinfo{author}{\bibfnamefont{K.-I.} \bibnamefont{Uchida}},
  \bibinfo{journal}{Proc. Jpn. Acad., Ser. B} \textbf{\bibinfo{volume}{97}},
  \bibinfo{pages}{69} (\bibinfo{year}{2021}).

\bibitem[{\citenamefont{Kikkawa and Saitoh}(2022)}]{kikkawa2022spin}
\bibinfo{author}{\bibfnamefont{T.}~\bibnamefont{Kikkawa}} \bibnamefont{and}
  \bibinfo{author}{\bibfnamefont{E.}~\bibnamefont{Saitoh}},
  \bibinfo{journal}{arXiv preprint arXiv:2205.10509}  (\bibinfo{year}{2022}).

\bibitem[{\citenamefont{Azevedo et~al.}(2005)\citenamefont{Azevedo,
  Vilela~Le{\~a}o, Rodriguez-Suarez, Oliveira, and Rezende}}]{azevedo2005dc}
\bibinfo{author}{\bibfnamefont{A.}~\bibnamefont{Azevedo}},
  \bibinfo{author}{\bibfnamefont{L.~H.} \bibnamefont{Vilela~Le{\~a}o}},
  \bibinfo{author}{\bibfnamefont{R.~L.} \bibnamefont{Rodriguez-Suarez}},
  \bibinfo{author}{\bibfnamefont{A.~B.} \bibnamefont{Oliveira}},
  \bibnamefont{and} \bibinfo{author}{\bibfnamefont{S.~M.}
  \bibnamefont{Rezende}}, \bibinfo{journal}{J. Appl. Phys.}
  \textbf{\bibinfo{volume}{97}}, \bibinfo{pages}{10C715}
  (\bibinfo{year}{2005}).

\bibitem[{\citenamefont{Saitoh et~al.}(2006)\citenamefont{Saitoh, Ueda,
  Miyajima, and Tatara}}]{saitoh2006conversion}
\bibinfo{author}{\bibfnamefont{E.}~\bibnamefont{Saitoh}},
  \bibinfo{author}{\bibfnamefont{M.}~\bibnamefont{Ueda}},
  \bibinfo{author}{\bibfnamefont{H.}~\bibnamefont{Miyajima}}, \bibnamefont{and}
  \bibinfo{author}{\bibfnamefont{G.}~\bibnamefont{Tatara}},
  \bibinfo{journal}{Appl. Phys. Lett.} \textbf{\bibinfo{volume}{88}},
  \bibinfo{pages}{182509} (\bibinfo{year}{2006}).

\bibitem[{\citenamefont{Kimura et~al.}(2007)\citenamefont{Kimura, Otani, Sato,
  Takahashi, and Maekawa}}]{kimura2007room}
\bibinfo{author}{\bibfnamefont{T.}~\bibnamefont{Kimura}},
  \bibinfo{author}{\bibfnamefont{Y.}~\bibnamefont{Otani}},
  \bibinfo{author}{\bibfnamefont{T.}~\bibnamefont{Sato}},
  \bibinfo{author}{\bibfnamefont{S.}~\bibnamefont{Takahashi}},
  \bibnamefont{and} \bibinfo{author}{\bibfnamefont{S.}~\bibnamefont{Maekawa}},
  \bibinfo{journal}{Phys. Rev. Lett.} \textbf{\bibinfo{volume}{98}},
  \bibinfo{pages}{156601} (\bibinfo{year}{2007}).

\bibitem[{\citenamefont{Valenzuela and Tinkham}(2006)}]{valenzuela2006direct}
\bibinfo{author}{\bibfnamefont{S.~O.} \bibnamefont{Valenzuela}}
  \bibnamefont{and} \bibinfo{author}{\bibfnamefont{M.}~\bibnamefont{Tinkham}},
  \bibinfo{journal}{Nature (London)} \textbf{\bibinfo{volume}{442}},
  \bibinfo{pages}{176} (\bibinfo{year}{2006}).

\bibitem[{\citenamefont{Sinova et~al.}(2015)\citenamefont{Sinova, Valenzuela,
  Wunderlich, Back, and Jungwirth}}]{sinova2015spin}
\bibinfo{author}{\bibfnamefont{J.}~\bibnamefont{Sinova}},
  \bibinfo{author}{\bibfnamefont{S.~O.} \bibnamefont{Valenzuela}},
  \bibinfo{author}{\bibfnamefont{J.}~\bibnamefont{Wunderlich}},
  \bibinfo{author}{\bibfnamefont{C.}~\bibnamefont{Back}}, \bibnamefont{and}
  \bibinfo{author}{\bibfnamefont{T.}~\bibnamefont{Jungwirth}},
  \bibinfo{journal}{Rev. Mod. Phys.} \textbf{\bibinfo{volume}{87}},
  \bibinfo{pages}{1213} (\bibinfo{year}{2015}).

\bibitem[{\citenamefont{Uchida et~al.}(2010)\citenamefont{Uchida, Xiao, Adachi,
  Ohe, Takahashi, Ieda, Ota, Kajiwara, Umezawa, Kawai et~al.}}]{uchida2010spin}
\bibinfo{author}{\bibfnamefont{K.}~\bibnamefont{Uchida}},
  \bibinfo{author}{\bibfnamefont{J.}~\bibnamefont{Xiao}},
  \bibinfo{author}{\bibfnamefont{H.}~\bibnamefont{Adachi}},
  \bibinfo{author}{\bibfnamefont{J.-i.} \bibnamefont{Ohe}},
  \bibinfo{author}{\bibfnamefont{S.}~\bibnamefont{Takahashi}},
  \bibinfo{author}{\bibfnamefont{J.}~\bibnamefont{Ieda}},
  \bibinfo{author}{\bibfnamefont{T.}~\bibnamefont{Ota}},
  \bibinfo{author}{\bibfnamefont{Y.}~\bibnamefont{Kajiwara}},
  \bibinfo{author}{\bibfnamefont{H.}~\bibnamefont{Umezawa}},
  \bibinfo{author}{\bibfnamefont{H.}~\bibnamefont{Kawai}},
  \bibnamefont{et~al.}, \bibinfo{journal}{Nat. Mater.}
  \textbf{\bibinfo{volume}{9}}, \bibinfo{pages}{894} (\bibinfo{year}{2010}).

\bibitem[{\citenamefont{Uchida et~al.}(2014{\natexlab{b}})\citenamefont{Uchida,
  Ishida, Kikkawa, Kirihara, Murakami, and Saitoh}}]{uchida2014longitudinal}
\bibinfo{author}{\bibfnamefont{K.}~\bibnamefont{Uchida}},
  \bibinfo{author}{\bibfnamefont{M.}~\bibnamefont{Ishida}},
  \bibinfo{author}{\bibfnamefont{T.}~\bibnamefont{Kikkawa}},
  \bibinfo{author}{\bibfnamefont{A.}~\bibnamefont{Kirihara}},
  \bibinfo{author}{\bibfnamefont{T.}~\bibnamefont{Murakami}}, \bibnamefont{and}
  \bibinfo{author}{\bibfnamefont{E.}~\bibnamefont{Saitoh}},
  \bibinfo{journal}{J. Phys. Condens. Matter} \textbf{\bibinfo{volume}{26}},
  \bibinfo{pages}{343202} (\bibinfo{year}{2014}{\natexlab{b}}).

\bibitem[{\citenamefont{Gepr{\"a}gs et~al.}(2016)\citenamefont{Gepr{\"a}gs,
  Kehlberger, Della~Coletta, Qiu, Guo, Schulz, Mix, Meyer, Kamra, Althammer
  et~al.}}]{geprags2016origin}
\bibinfo{author}{\bibfnamefont{S.}~\bibnamefont{Gepr{\"a}gs}},
  \bibinfo{author}{\bibfnamefont{A.}~\bibnamefont{Kehlberger}},
  \bibinfo{author}{\bibfnamefont{F.}~\bibnamefont{Della~Coletta}},
  \bibinfo{author}{\bibfnamefont{Z.}~\bibnamefont{Qiu}},
  \bibinfo{author}{\bibfnamefont{E.-J.} \bibnamefont{Guo}},
  \bibinfo{author}{\bibfnamefont{T.}~\bibnamefont{Schulz}},
  \bibinfo{author}{\bibfnamefont{C.}~\bibnamefont{Mix}},
  \bibinfo{author}{\bibfnamefont{S.}~\bibnamefont{Meyer}},
  \bibinfo{author}{\bibfnamefont{A.}~\bibnamefont{Kamra}},
  \bibinfo{author}{\bibfnamefont{M.}~\bibnamefont{Althammer}},
  \bibnamefont{et~al.}, \bibinfo{journal}{Nat. Commun.}
  \textbf{\bibinfo{volume}{7}}, \bibinfo{pages}{10452} (\bibinfo{year}{2016}).

\bibitem[{\citenamefont{Oh et~al.}(2021)\citenamefont{Oh, Park, Choe, Jo,
  Jeong, Jin, Jo, Suh, Min, and Yoo}}]{oh2021scalable}
\bibinfo{author}{\bibfnamefont{I.}~\bibnamefont{Oh}},
  \bibinfo{author}{\bibfnamefont{J.}~\bibnamefont{Park}},
  \bibinfo{author}{\bibfnamefont{D.}~\bibnamefont{Choe}},
  \bibinfo{author}{\bibfnamefont{J.}~\bibnamefont{Jo}},
  \bibinfo{author}{\bibfnamefont{H.}~\bibnamefont{Jeong}},
  \bibinfo{author}{\bibfnamefont{M.-J.} \bibnamefont{Jin}},
  \bibinfo{author}{\bibfnamefont{Y.}~\bibnamefont{Jo}},
  \bibinfo{author}{\bibfnamefont{J.}~\bibnamefont{Suh}},
  \bibinfo{author}{\bibfnamefont{B.-C.} \bibnamefont{Min}}, \bibnamefont{and}
  \bibinfo{author}{\bibfnamefont{J.-W.} \bibnamefont{Yoo}},
  \bibinfo{journal}{Nat. Commun.} \textbf{\bibinfo{volume}{12}},
  \bibinfo{pages}{1057} (\bibinfo{year}{2021}).

\bibitem[{\citenamefont{Ito et~al.}(2019)\citenamefont{Ito, Kikkawa, Barker,
  Hirobe, Shiomi, and Saitoh}}]{ito2019spin}
\bibinfo{author}{\bibfnamefont{N.}~\bibnamefont{Ito}},
  \bibinfo{author}{\bibfnamefont{T.}~\bibnamefont{Kikkawa}},
  \bibinfo{author}{\bibfnamefont{J.}~\bibnamefont{Barker}},
  \bibinfo{author}{\bibfnamefont{D.}~\bibnamefont{Hirobe}},
  \bibinfo{author}{\bibfnamefont{Y.}~\bibnamefont{Shiomi}}, \bibnamefont{and}
  \bibinfo{author}{\bibfnamefont{E.}~\bibnamefont{Saitoh}},
  \bibinfo{journal}{Phys. Rev. B} \textbf{\bibinfo{volume}{100}},
  \bibinfo{pages}{060402} (\bibinfo{year}{2019}).

\bibitem[{\citenamefont{Mallick et~al.}(2019)\citenamefont{Mallick, Wagh,
  Ionescu, Barnes, and Kumar}}]{mallick2019role}
\bibinfo{author}{\bibfnamefont{K.}~\bibnamefont{Mallick}},
  \bibinfo{author}{\bibfnamefont{A.~A.} \bibnamefont{Wagh}},
  \bibinfo{author}{\bibfnamefont{A.}~\bibnamefont{Ionescu}},
  \bibinfo{author}{\bibfnamefont{C.~H.} \bibnamefont{Barnes}},
  \bibnamefont{and} \bibinfo{author}{\bibfnamefont{P.~A.} \bibnamefont{Kumar}},
  \bibinfo{journal}{Phys. Rev. B} \textbf{\bibinfo{volume}{100}},
  \bibinfo{pages}{224403} (\bibinfo{year}{2019}).

\bibitem[{\citenamefont{Seki et~al.}(2015)\citenamefont{Seki, Ideue, Kubota,
  Kozuka, Takagi, Nakamura, Kaneko, Kawasaki, and Tokura}}]{seki2015thermal}
\bibinfo{author}{\bibfnamefont{S.}~\bibnamefont{Seki}},
  \bibinfo{author}{\bibfnamefont{T.}~\bibnamefont{Ideue}},
  \bibinfo{author}{\bibfnamefont{M.}~\bibnamefont{Kubota}},
  \bibinfo{author}{\bibfnamefont{Y.}~\bibnamefont{Kozuka}},
  \bibinfo{author}{\bibfnamefont{R.}~\bibnamefont{Takagi}},
  \bibinfo{author}{\bibfnamefont{M.}~\bibnamefont{Nakamura}},
  \bibinfo{author}{\bibfnamefont{Y.}~\bibnamefont{Kaneko}},
  \bibinfo{author}{\bibfnamefont{M.}~\bibnamefont{Kawasaki}}, \bibnamefont{and}
  \bibinfo{author}{\bibfnamefont{Y.}~\bibnamefont{Tokura}},
  \bibinfo{journal}{Phys. Rev. Lett.} \textbf{\bibinfo{volume}{115}},
  \bibinfo{pages}{266601} (\bibinfo{year}{2015}).

\bibitem[{\citenamefont{Wu et~al.}(2016)\citenamefont{Wu, Zhang, Amit, Borisov,
  Pearson, Jiang, Lederman, Hoffmann, and
  Bhattacharya}}]{wu2016antiferromagnetic}
\bibinfo{author}{\bibfnamefont{S.~M.} \bibnamefont{Wu}},
  \bibinfo{author}{\bibfnamefont{W.}~\bibnamefont{Zhang}},
  \bibinfo{author}{\bibfnamefont{K.~C.} \bibnamefont{Amit}},
  \bibinfo{author}{\bibfnamefont{P.}~\bibnamefont{Borisov}},
  \bibinfo{author}{\bibfnamefont{J.~E.} \bibnamefont{Pearson}},
  \bibinfo{author}{\bibfnamefont{J.~S.} \bibnamefont{Jiang}},
  \bibinfo{author}{\bibfnamefont{D.}~\bibnamefont{Lederman}},
  \bibinfo{author}{\bibfnamefont{A.}~\bibnamefont{Hoffmann}}, \bibnamefont{and}
  \bibinfo{author}{\bibfnamefont{A.}~\bibnamefont{Bhattacharya}},
  \bibinfo{journal}{Phys. Rev. Lett.} \textbf{\bibinfo{volume}{116}},
  \bibinfo{pages}{097204} (\bibinfo{year}{2016}).

\bibitem[{\citenamefont{Rezende et~al.}(2016)\citenamefont{Rezende,
  Rodr{\'\i}guez-Su{\'a}rez, and Azevedo}}]{rezende2016theory}
\bibinfo{author}{\bibfnamefont{S.}~\bibnamefont{Rezende}},
  \bibinfo{author}{\bibfnamefont{R.}~\bibnamefont{Rodr{\'\i}guez-Su{\'a}rez}},
  \bibnamefont{and} \bibinfo{author}{\bibfnamefont{A.}~\bibnamefont{Azevedo}},
  \bibinfo{journal}{Phys. Rev. B} \textbf{\bibinfo{volume}{93}},
  \bibinfo{pages}{014425} (\bibinfo{year}{2016}).

\bibitem[{\citenamefont{Li et~al.}(2020)\citenamefont{Li, Wilson, Cheng,
  Lohmann, Kavand, Yuan, Aldosary, Agladze, Wei, Sherwin et~al.}}]{li2020spin}
\bibinfo{author}{\bibfnamefont{J.}~\bibnamefont{Li}},
  \bibinfo{author}{\bibfnamefont{C.~B.} \bibnamefont{Wilson}},
  \bibinfo{author}{\bibfnamefont{R.}~\bibnamefont{Cheng}},
  \bibinfo{author}{\bibfnamefont{M.}~\bibnamefont{Lohmann}},
  \bibinfo{author}{\bibfnamefont{M.}~\bibnamefont{Kavand}},
  \bibinfo{author}{\bibfnamefont{W.}~\bibnamefont{Yuan}},
  \bibinfo{author}{\bibfnamefont{M.}~\bibnamefont{Aldosary}},
  \bibinfo{author}{\bibfnamefont{N.}~\bibnamefont{Agladze}},
  \bibinfo{author}{\bibfnamefont{P.}~\bibnamefont{Wei}},
  \bibinfo{author}{\bibfnamefont{M.~S.} \bibnamefont{Sherwin}},
  \bibnamefont{et~al.}, \bibinfo{journal}{Nature (London)}
  \textbf{\bibinfo{volume}{578}}, \bibinfo{pages}{70} (\bibinfo{year}{2020}).

\bibitem[{\citenamefont{Reitz et~al.}(2020)\citenamefont{Reitz, Li, Yuan, Shi,
  and Tserkovnyak}}]{reitz2020spin}
\bibinfo{author}{\bibfnamefont{D.}~\bibnamefont{Reitz}},
  \bibinfo{author}{\bibfnamefont{J.}~\bibnamefont{Li}},
  \bibinfo{author}{\bibfnamefont{W.}~\bibnamefont{Yuan}},
  \bibinfo{author}{\bibfnamefont{J.}~\bibnamefont{Shi}}, \bibnamefont{and}
  \bibinfo{author}{\bibfnamefont{Y.}~\bibnamefont{Tserkovnyak}},
  \bibinfo{journal}{Phys. Rev. B} \textbf{\bibinfo{volume}{102}},
  \bibinfo{pages}{020408} (\bibinfo{year}{2020}).

\bibitem[{\citenamefont{Kikkawa et~al.}(2021)\citenamefont{Kikkawa, Reitz, Ito,
  Makiuchi, Sugimoto, Tsunekawa, Daimon, Oyanagi, Ramos, Takahashi
  et~al.}}]{kikkawa2021observation}
\bibinfo{author}{\bibfnamefont{T.}~\bibnamefont{Kikkawa}},
  \bibinfo{author}{\bibfnamefont{D.}~\bibnamefont{Reitz}},
  \bibinfo{author}{\bibfnamefont{H.}~\bibnamefont{Ito}},
  \bibinfo{author}{\bibfnamefont{T.}~\bibnamefont{Makiuchi}},
  \bibinfo{author}{\bibfnamefont{T.}~\bibnamefont{Sugimoto}},
  \bibinfo{author}{\bibfnamefont{K.}~\bibnamefont{Tsunekawa}},
  \bibinfo{author}{\bibfnamefont{S.}~\bibnamefont{Daimon}},
  \bibinfo{author}{\bibfnamefont{K.}~\bibnamefont{Oyanagi}},
  \bibinfo{author}{\bibfnamefont{R.}~\bibnamefont{Ramos}},
  \bibinfo{author}{\bibfnamefont{S.}~\bibnamefont{Takahashi}},
  \bibnamefont{et~al.}, \bibinfo{journal}{Nat. Commun.}
  \textbf{\bibinfo{volume}{12}}, \bibinfo{pages}{4356} (\bibinfo{year}{2021}).

\bibitem[{\citenamefont{Xiao et~al.}(2010)\citenamefont{Xiao, Bauer, Uchida,
  Saitoh, Maekawa et~al.}}]{xiao2010theory}
\bibinfo{author}{\bibfnamefont{J.}~\bibnamefont{Xiao}},
  \bibinfo{author}{\bibfnamefont{G.~E.} \bibnamefont{Bauer}},
  \bibinfo{author}{\bibfnamefont{K.-c.} \bibnamefont{Uchida}},
  \bibinfo{author}{\bibfnamefont{E.}~\bibnamefont{Saitoh}},
  \bibinfo{author}{\bibfnamefont{S.}~\bibnamefont{Maekawa}},
  \bibnamefont{et~al.}, \bibinfo{journal}{Phys. Rev. B}
  \textbf{\bibinfo{volume}{81}}, \bibinfo{pages}{214418}
  (\bibinfo{year}{2010}).

\bibitem[{\citenamefont{Bender et~al.}(2012)\citenamefont{Bender, Duine, and
  Tserkovnyak}}]{bender2012electronic}
\bibinfo{author}{\bibfnamefont{S.~A.} \bibnamefont{Bender}},
  \bibinfo{author}{\bibfnamefont{R.~A.} \bibnamefont{Duine}}, \bibnamefont{and}
  \bibinfo{author}{\bibfnamefont{Y.}~\bibnamefont{Tserkovnyak}},
  \bibinfo{journal}{Phys. Rev. Lett.} \textbf{\bibinfo{volume}{108}},
  \bibinfo{pages}{246601} (\bibinfo{year}{2012}).

\bibitem[{\citenamefont{Adachi et~al.}(2013)\citenamefont{Adachi, Uchida,
  Saitoh, and Maekawa}}]{adachi2013theory}
\bibinfo{author}{\bibfnamefont{H.}~\bibnamefont{Adachi}},
  \bibinfo{author}{\bibfnamefont{K.-i.} \bibnamefont{Uchida}},
  \bibinfo{author}{\bibfnamefont{E.}~\bibnamefont{Saitoh}}, \bibnamefont{and}
  \bibinfo{author}{\bibfnamefont{S.}~\bibnamefont{Maekawa}},
  \bibinfo{journal}{Rep. Prog. Phys.} \textbf{\bibinfo{volume}{76}},
  \bibinfo{pages}{036501} (\bibinfo{year}{2013}).

\bibitem[{\citenamefont{Rezende et~al.}(2014)\citenamefont{Rezende,
  Rodr{\'\i}guez-Su{\'a}rez, Cunha, Rodrigues, Machado, Guerra, Ortiz, and
  Azevedo}}]{rezende2014magnon}
\bibinfo{author}{\bibfnamefont{S.~M.} \bibnamefont{Rezende}},
  \bibinfo{author}{\bibfnamefont{R.~L.}
  \bibnamefont{Rodr{\'\i}guez-Su{\'a}rez}},
  \bibinfo{author}{\bibfnamefont{R.}~\bibnamefont{Cunha}},
  \bibinfo{author}{\bibfnamefont{A.~R.} \bibnamefont{Rodrigues}},
  \bibinfo{author}{\bibfnamefont{F.~L.~A.} \bibnamefont{Machado}},
  \bibinfo{author}{\bibfnamefont{G.~A.~F.} \bibnamefont{Guerra}},
  \bibinfo{author}{\bibfnamefont{J.~C.~L.} \bibnamefont{Ortiz}},
  \bibnamefont{and} \bibinfo{author}{\bibfnamefont{A.}~\bibnamefont{Azevedo}},
  \bibinfo{journal}{Phys. Rev. B} \textbf{\bibinfo{volume}{89}},
  \bibinfo{pages}{014416} (\bibinfo{year}{2014}).

\bibitem[{\citenamefont{Rezende}(2020)}]{rezende2020fundamentals}
\bibinfo{author}{\bibfnamefont{S.~M.} \bibnamefont{Rezende}},
  \emph{\bibinfo{title}{Fundamentals of magnonics}}, vol. \bibinfo{volume}{969}
  (\bibinfo{publisher}{Springer}, \bibinfo{year}{2020}).

\bibitem[{\citenamefont{Wu et~al.}(2015{\natexlab{a}})\citenamefont{Wu,
  Pearson, and Bhattacharya}}]{wu2015paramagnetic}
\bibinfo{author}{\bibfnamefont{S.~M.} \bibnamefont{Wu}},
  \bibinfo{author}{\bibfnamefont{J.~E.} \bibnamefont{Pearson}},
  \bibnamefont{and}
  \bibinfo{author}{\bibfnamefont{A.}~\bibnamefont{Bhattacharya}},
  \bibinfo{journal}{Phys. Rev. Lett.} \textbf{\bibinfo{volume}{114}},
  \bibinfo{pages}{186602} (\bibinfo{year}{2015}{\natexlab{a}}).

\bibitem[{\citenamefont{Aqeel et~al.}(2015)\citenamefont{Aqeel, Vlietstra,
  Heuver, Bauer, Noheda, van Wees, and Palstra}}]{aqeel2015spin}
\bibinfo{author}{\bibfnamefont{A.}~\bibnamefont{Aqeel}},
  \bibinfo{author}{\bibfnamefont{N.}~\bibnamefont{Vlietstra}},
  \bibinfo{author}{\bibfnamefont{J.~A.} \bibnamefont{Heuver}},
  \bibinfo{author}{\bibfnamefont{G.~E.~W.} \bibnamefont{Bauer}},
  \bibinfo{author}{\bibfnamefont{B.}~\bibnamefont{Noheda}},
  \bibinfo{author}{\bibfnamefont{B.~J.} \bibnamefont{van Wees}},
  \bibnamefont{and} \bibinfo{author}{\bibfnamefont{T.~T.~M.}
  \bibnamefont{Palstra}}, \bibinfo{journal}{Phys. Rev. B}
  \textbf{\bibinfo{volume}{92}}, \bibinfo{pages}{224410}
  (\bibinfo{year}{2015}).

\bibitem[{\citenamefont{Li et~al.}(2019)\citenamefont{Li, Shi, Ortiz, Aldosary,
  Chen, Aji, Wei, and Shi}}]{li2019spin}
\bibinfo{author}{\bibfnamefont{J.}~\bibnamefont{Li}},
  \bibinfo{author}{\bibfnamefont{Z.}~\bibnamefont{Shi}},
  \bibinfo{author}{\bibfnamefont{V.~H.} \bibnamefont{Ortiz}},
  \bibinfo{author}{\bibfnamefont{M.}~\bibnamefont{Aldosary}},
  \bibinfo{author}{\bibfnamefont{C.}~\bibnamefont{Chen}},
  \bibinfo{author}{\bibfnamefont{V.}~\bibnamefont{Aji}},
  \bibinfo{author}{\bibfnamefont{P.}~\bibnamefont{Wei}}, \bibnamefont{and}
  \bibinfo{author}{\bibfnamefont{J.}~\bibnamefont{Shi}},
  \bibinfo{journal}{Phys. Rev. Lett.} \textbf{\bibinfo{volume}{122}},
  \bibinfo{pages}{217204} (\bibinfo{year}{2019}).

\bibitem[{\citenamefont{Hong et~al.}(2019)\citenamefont{Hong, Liu, Pearson,
  Hoffmann, Fong, and Bhattacharya}}]{hong2019spin}
\bibinfo{author}{\bibfnamefont{D.}~\bibnamefont{Hong}},
  \bibinfo{author}{\bibfnamefont{C.}~\bibnamefont{Liu}},
  \bibinfo{author}{\bibfnamefont{J.~E.} \bibnamefont{Pearson}},
  \bibinfo{author}{\bibfnamefont{A.}~\bibnamefont{Hoffmann}},
  \bibinfo{author}{\bibfnamefont{D.~D.} \bibnamefont{Fong}}, \bibnamefont{and}
  \bibinfo{author}{\bibfnamefont{A.}~\bibnamefont{Bhattacharya}},
  \bibinfo{journal}{Appl. Phys. Lett.} \textbf{\bibinfo{volume}{114}},
  \bibinfo{pages}{242403} (\bibinfo{year}{2019}).

\bibitem[{\citenamefont{Hirobe et~al.}(2017)\citenamefont{Hirobe, Sato,
  Kawamata, Shiomi, Uchida, Iguchi, Koike, Maekawa, and
  Saitoh}}]{hirobe2017one}
\bibinfo{author}{\bibfnamefont{D.}~\bibnamefont{Hirobe}},
  \bibinfo{author}{\bibfnamefont{M.}~\bibnamefont{Sato}},
  \bibinfo{author}{\bibfnamefont{T.}~\bibnamefont{Kawamata}},
  \bibinfo{author}{\bibfnamefont{Y.}~\bibnamefont{Shiomi}},
  \bibinfo{author}{\bibfnamefont{K.-i.} \bibnamefont{Uchida}},
  \bibinfo{author}{\bibfnamefont{R.}~\bibnamefont{Iguchi}},
  \bibinfo{author}{\bibfnamefont{Y.}~\bibnamefont{Koike}},
  \bibinfo{author}{\bibfnamefont{S.}~\bibnamefont{Maekawa}}, \bibnamefont{and}
  \bibinfo{author}{\bibfnamefont{E.}~\bibnamefont{Saitoh}},
  \bibinfo{journal}{Nat. Phys.} \textbf{\bibinfo{volume}{13}},
  \bibinfo{pages}{30} (\bibinfo{year}{2017}).

\bibitem[{\citenamefont{Hirobe et~al.}(2018)\citenamefont{Hirobe, Kawamata,
  Oyanagi, Koike, and Saitoh}}]{hirobe2018generation}
\bibinfo{author}{\bibfnamefont{D.}~\bibnamefont{Hirobe}},
  \bibinfo{author}{\bibfnamefont{T.}~\bibnamefont{Kawamata}},
  \bibinfo{author}{\bibfnamefont{K.}~\bibnamefont{Oyanagi}},
  \bibinfo{author}{\bibfnamefont{Y.}~\bibnamefont{Koike}}, \bibnamefont{and}
  \bibinfo{author}{\bibfnamefont{E.}~\bibnamefont{Saitoh}},
  \bibinfo{journal}{J. Appl. Phys.} \textbf{\bibinfo{volume}{123}},
  \bibinfo{pages}{123903} (\bibinfo{year}{2018}).

\bibitem[{\citenamefont{Chen et~al.}(2021)\citenamefont{Chen, Sato, Tang,
  Shiomi, Oyanagi, Masuda, Nambu, Fujita, and Saitoh}}]{chen2021triplon}
\bibinfo{author}{\bibfnamefont{Y.}~\bibnamefont{Chen}},
  \bibinfo{author}{\bibfnamefont{M.}~\bibnamefont{Sato}},
  \bibinfo{author}{\bibfnamefont{Y.}~\bibnamefont{Tang}},
  \bibinfo{author}{\bibfnamefont{Y.}~\bibnamefont{Shiomi}},
  \bibinfo{author}{\bibfnamefont{K.}~\bibnamefont{Oyanagi}},
  \bibinfo{author}{\bibfnamefont{T.}~\bibnamefont{Masuda}},
  \bibinfo{author}{\bibfnamefont{Y.}~\bibnamefont{Nambu}},
  \bibinfo{author}{\bibfnamefont{M.}~\bibnamefont{Fujita}}, \bibnamefont{and}
  \bibinfo{author}{\bibfnamefont{E.}~\bibnamefont{Saitoh}},
  \bibinfo{journal}{Nat. Commun.} \textbf{\bibinfo{volume}{12}},
  \bibinfo{pages}{5199} (\bibinfo{year}{2021}).

\bibitem[{\citenamefont{Xing et~al.}(2022)\citenamefont{Xing, Cai, Moriyama,
  Nara, Yao, Qiao, Yoshimura, and Han}}]{xing2022spin}
\bibinfo{author}{\bibfnamefont{W.}~\bibnamefont{Xing}},
  \bibinfo{author}{\bibfnamefont{R.}~\bibnamefont{Cai}},
  \bibinfo{author}{\bibfnamefont{K.}~\bibnamefont{Moriyama}},
  \bibinfo{author}{\bibfnamefont{K.}~\bibnamefont{Nara}},
  \bibinfo{author}{\bibfnamefont{Y.}~\bibnamefont{Yao}},
  \bibinfo{author}{\bibfnamefont{W.}~\bibnamefont{Qiao}},
  \bibinfo{author}{\bibfnamefont{K.}~\bibnamefont{Yoshimura}},
  \bibnamefont{and} \bibinfo{author}{\bibfnamefont{W.}~\bibnamefont{Han}},
  \bibinfo{journal}{Appl. Phys. Lett.} \textbf{\bibinfo{volume}{120}},
  \bibinfo{pages}{042402} (\bibinfo{year}{2022}).

\bibitem[{\citenamefont{Luo et~al.}(2022)\citenamefont{Luo, Zhao, Chen,
  Legvold, Navarro, Schuller, and Natelson}}]{luo2022spin}
\bibinfo{author}{\bibfnamefont{R.}~\bibnamefont{Luo}},
  \bibinfo{author}{\bibfnamefont{X.}~\bibnamefont{Zhao}},
  \bibinfo{author}{\bibfnamefont{L.}~\bibnamefont{Chen}},
  \bibinfo{author}{\bibfnamefont{T.~J.} \bibnamefont{Legvold}},
  \bibinfo{author}{\bibfnamefont{H.}~\bibnamefont{Navarro}},
  \bibinfo{author}{\bibfnamefont{I.~K.} \bibnamefont{Schuller}},
  \bibnamefont{and} \bibinfo{author}{\bibfnamefont{D.}~\bibnamefont{Natelson}},
  \bibinfo{journal}{Appl. Phys. Lett.} \textbf{\bibinfo{volume}{121}},
  \bibinfo{pages}{102404} (\bibinfo{year}{2022}).

\bibitem[{\citenamefont{Okamoto}(2016)}]{okamoto2016spin}
\bibinfo{author}{\bibfnamefont{S.}~\bibnamefont{Okamoto}},
  \bibinfo{journal}{Phys. Rev. B} \textbf{\bibinfo{volume}{93}},
  \bibinfo{pages}{064421} (\bibinfo{year}{2016}).

\bibitem[{\citenamefont{Yamamoto et~al.}(2019)\citenamefont{Yamamoto, Ichioka,
  and Adachi}}]{yamamoto2019spin}
\bibinfo{author}{\bibfnamefont{Y.}~\bibnamefont{Yamamoto}},
  \bibinfo{author}{\bibfnamefont{M.}~\bibnamefont{Ichioka}}, \bibnamefont{and}
  \bibinfo{author}{\bibfnamefont{H.}~\bibnamefont{Adachi}},
  \bibinfo{journal}{Phys. Rev. B} \textbf{\bibinfo{volume}{100}},
  \bibinfo{pages}{064419} (\bibinfo{year}{2019}).

\bibitem[{\citenamefont{Zhang et~al.}(2019)\citenamefont{Zhang, Bergeret, and
  Golovach}}]{zhang2019theory}
\bibinfo{author}{\bibfnamefont{X.-P.} \bibnamefont{Zhang}},
  \bibinfo{author}{\bibfnamefont{F.~S.} \bibnamefont{Bergeret}},
  \bibnamefont{and} \bibinfo{author}{\bibfnamefont{V.~N.}
  \bibnamefont{Golovach}}, \bibinfo{journal}{Nano Lett.}
  \textbf{\bibinfo{volume}{19}}, \bibinfo{pages}{6330} (\bibinfo{year}{2019}).

\bibitem[{\citenamefont{Nakata and Ohnuma}(2021)}]{nakata2021magnonic}
\bibinfo{author}{\bibfnamefont{K.}~\bibnamefont{Nakata}} \bibnamefont{and}
  \bibinfo{author}{\bibfnamefont{Y.}~\bibnamefont{Ohnuma}},
  \bibinfo{journal}{Phys. Rev. B} \textbf{\bibinfo{volume}{104}},
  \bibinfo{pages}{064408} (\bibinfo{year}{2021}).

\bibitem[{\citenamefont{Kajiwara et~al.}(2010)\citenamefont{Kajiwara, Harii,
  Takahashi, Ohe, Uchida, Mizuguchi, Umezawa, Kawai, Ando, Takanashi
  et~al.}}]{kajiwara2010transmission}
\bibinfo{author}{\bibfnamefont{Y.}~\bibnamefont{Kajiwara}},
  \bibinfo{author}{\bibfnamefont{K.}~\bibnamefont{Harii}},
  \bibinfo{author}{\bibfnamefont{S.}~\bibnamefont{Takahashi}},
  \bibinfo{author}{\bibfnamefont{J.-i.} \bibnamefont{Ohe}},
  \bibinfo{author}{\bibfnamefont{K.}~\bibnamefont{Uchida}},
  \bibinfo{author}{\bibfnamefont{M.}~\bibnamefont{Mizuguchi}},
  \bibinfo{author}{\bibfnamefont{H.}~\bibnamefont{Umezawa}},
  \bibinfo{author}{\bibfnamefont{H.}~\bibnamefont{Kawai}},
  \bibinfo{author}{\bibfnamefont{K.}~\bibnamefont{Ando}},
  \bibinfo{author}{\bibfnamefont{K.}~\bibnamefont{Takanashi}},
  \bibnamefont{et~al.}, \bibinfo{journal}{Nature (London)}
  \textbf{\bibinfo{volume}{464}}, \bibinfo{pages}{262} (\bibinfo{year}{2010}).

\bibitem[{\citenamefont{Takahashi et~al.}(2010)\citenamefont{Takahashi, Saitoh,
  and Maekawa}}]{takahashi2010spin}
\bibinfo{author}{\bibfnamefont{S.}~\bibnamefont{Takahashi}},
  \bibinfo{author}{\bibfnamefont{E.}~\bibnamefont{Saitoh}}, \bibnamefont{and}
  \bibinfo{author}{\bibfnamefont{S.}~\bibnamefont{Maekawa}}, in
  \emph{\bibinfo{booktitle}{J. Phys.: Conf. Ser.}} (\bibinfo{organization}{IOP
  Publishing}, \bibinfo{year}{2010}), vol. \bibinfo{volume}{200}, p.
  \bibinfo{pages}{062030}.

\bibitem[{\citenamefont{Cornelissen et~al.}(2016)\citenamefont{Cornelissen,
  Peters, Bauer, Duine, and van Wees}}]{cornelissen2016magnon}
\bibinfo{author}{\bibfnamefont{L.~J.} \bibnamefont{Cornelissen}},
  \bibinfo{author}{\bibfnamefont{K.~J.~H.} \bibnamefont{Peters}},
  \bibinfo{author}{\bibfnamefont{G.~E.~W.} \bibnamefont{Bauer}},
  \bibinfo{author}{\bibfnamefont{R.~A.} \bibnamefont{Duine}}, \bibnamefont{and}
  \bibinfo{author}{\bibfnamefont{B.~J.} \bibnamefont{van Wees}},
  \bibinfo{journal}{Phys. Rev. B} \textbf{\bibinfo{volume}{94}},
  \bibinfo{pages}{014412} (\bibinfo{year}{2016}).

\bibitem[{\citenamefont{Hu et~al.}(2014)\citenamefont{Hu, Zhang, Huang, and
  Xia}}]{hu2014effective}
\bibinfo{author}{\bibfnamefont{F.}~\bibnamefont{Hu}},
  \bibinfo{author}{\bibfnamefont{G.-Y.} \bibnamefont{Zhang}},
  \bibinfo{author}{\bibfnamefont{Y.-J.} \bibnamefont{Huang}}, \bibnamefont{and}
  \bibinfo{author}{\bibfnamefont{W.-S.} \bibnamefont{Xia}},
  \bibinfo{journal}{Chinese Physics Letters} \textbf{\bibinfo{volume}{31}},
  \bibinfo{pages}{057501} (\bibinfo{year}{2014}).

\bibitem[{\citenamefont{Daudin et~al.}(1982)\citenamefont{Daudin, Lagnier, and
  Salce}}]{daudin1982thermodynamic}
\bibinfo{author}{\bibfnamefont{B.}~\bibnamefont{Daudin}},
  \bibinfo{author}{\bibfnamefont{R.}~\bibnamefont{Lagnier}}, \bibnamefont{and}
  \bibinfo{author}{\bibfnamefont{B.}~\bibnamefont{Salce}}, \bibinfo{journal}{J.
  Mag. Mag. Mater.} \textbf{\bibinfo{volume}{27}}, \bibinfo{pages}{315}
  (\bibinfo{year}{1982}).

\bibitem[{\citenamefont{Oyanagi et~al.}(2019)\citenamefont{Oyanagi, Takahashi,
  Cornelissen, Shan, Daimon, Kikkawa, Bauer, van Wees, and
  Saitoh}}]{oyanagi2019spin}
\bibinfo{author}{\bibfnamefont{K.}~\bibnamefont{Oyanagi}},
  \bibinfo{author}{\bibfnamefont{S.}~\bibnamefont{Takahashi}},
  \bibinfo{author}{\bibfnamefont{L.~J.} \bibnamefont{Cornelissen}},
  \bibinfo{author}{\bibfnamefont{J.}~\bibnamefont{Shan}},
  \bibinfo{author}{\bibfnamefont{S.}~\bibnamefont{Daimon}},
  \bibinfo{author}{\bibfnamefont{T.}~\bibnamefont{Kikkawa}},
  \bibinfo{author}{\bibfnamefont{G.~E.~W.} \bibnamefont{Bauer}},
  \bibinfo{author}{\bibfnamefont{B.~J.} \bibnamefont{van Wees}},
  \bibnamefont{and} \bibinfo{author}{\bibfnamefont{E.}~\bibnamefont{Saitoh}},
  \bibinfo{journal}{Nat. Commun.} \textbf{\bibinfo{volume}{10}},
  \bibinfo{pages}{4740} (\bibinfo{year}{2019}).

\bibitem[{\citenamefont{Oyanagi et~al.}(2021)\citenamefont{Oyanagi,
  Gomez-Perez, Zhang, Kikkawa, Chen, Sagasta, Chuvilin, Hueso, Golovach,
  Bergeret et~al.}}]{oyanagi2021paramagnetic}
\bibinfo{author}{\bibfnamefont{K.}~\bibnamefont{Oyanagi}},
  \bibinfo{author}{\bibfnamefont{J.~M.} \bibnamefont{Gomez-Perez}},
  \bibinfo{author}{\bibfnamefont{X.-P.} \bibnamefont{Zhang}},
  \bibinfo{author}{\bibfnamefont{T.}~\bibnamefont{Kikkawa}},
  \bibinfo{author}{\bibfnamefont{Y.}~\bibnamefont{Chen}},
  \bibinfo{author}{\bibfnamefont{E.}~\bibnamefont{Sagasta}},
  \bibinfo{author}{\bibfnamefont{A.}~\bibnamefont{Chuvilin}},
  \bibinfo{author}{\bibfnamefont{L.~E.} \bibnamefont{Hueso}},
  \bibinfo{author}{\bibfnamefont{V.~N.} \bibnamefont{Golovach}},
  \bibinfo{author}{\bibfnamefont{F.~S.} \bibnamefont{Bergeret}},
  \bibnamefont{et~al.}, \bibinfo{journal}{Phys. Rev. B}
  \textbf{\bibinfo{volume}{104}}, \bibinfo{pages}{134428}
  (\bibinfo{year}{2021}).

\bibitem[{\citenamefont{Liu et~al.}(2018)\citenamefont{Liu, Wu, Pearson, Jiang,
  d'Ambrumenil, and Bhattacharya}}]{liu2018probing}
\bibinfo{author}{\bibfnamefont{C.}~\bibnamefont{Liu}},
  \bibinfo{author}{\bibfnamefont{S.~M.} \bibnamefont{Wu}},
  \bibinfo{author}{\bibfnamefont{J.~E.} \bibnamefont{Pearson}},
  \bibinfo{author}{\bibfnamefont{J.~S.} \bibnamefont{Jiang}},
  \bibinfo{author}{\bibfnamefont{N.}~\bibnamefont{d'Ambrumenil}},
  \bibnamefont{and}
  \bibinfo{author}{\bibfnamefont{A.}~\bibnamefont{Bhattacharya}},
  \bibinfo{journal}{Phys. Rev. B} \textbf{\bibinfo{volume}{98}},
  \bibinfo{pages}{060415} (\bibinfo{year}{2018}).

\bibitem[{\citenamefont{Sun et~al.}(2012)\citenamefont{Sun, Song, Chang,
  Kabatek, Jantz, Schneider, Wu, Schultheiss, and Hoffmann}}]{sun2012growth}
\bibinfo{author}{\bibfnamefont{Y.}~\bibnamefont{Sun}},
  \bibinfo{author}{\bibfnamefont{Y.-Y.} \bibnamefont{Song}},
  \bibinfo{author}{\bibfnamefont{H.}~\bibnamefont{Chang}},
  \bibinfo{author}{\bibfnamefont{M.}~\bibnamefont{Kabatek}},
  \bibinfo{author}{\bibfnamefont{M.}~\bibnamefont{Jantz}},
  \bibinfo{author}{\bibfnamefont{W.}~\bibnamefont{Schneider}},
  \bibinfo{author}{\bibfnamefont{M.}~\bibnamefont{Wu}},
  \bibinfo{author}{\bibfnamefont{H.}~\bibnamefont{Schultheiss}},
  \bibnamefont{and} \bibinfo{author}{\bibfnamefont{A.}~\bibnamefont{Hoffmann}},
  \bibinfo{journal}{Appl. Phys. Lett.} \textbf{\bibinfo{volume}{101}},
  \bibinfo{pages}{152405} (\bibinfo{year}{2012}).

\bibitem[{\citenamefont{Serga et~al.}(2010)\citenamefont{Serga, Chumak, and
  Hillebrands}}]{serga2010yig}
\bibinfo{author}{\bibfnamefont{A.~A.} \bibnamefont{Serga}},
  \bibinfo{author}{\bibfnamefont{A.~V.} \bibnamefont{Chumak}},
  \bibnamefont{and}
  \bibinfo{author}{\bibfnamefont{B.}~\bibnamefont{Hillebrands}},
  \bibinfo{journal}{J. Phys. D} \textbf{\bibinfo{volume}{43}},
  \bibinfo{pages}{264002} (\bibinfo{year}{2010}).

\bibitem[{\citenamefont{Chumak et~al.}(2015)\citenamefont{Chumak, Vasyuchka,
  Serga, and Hillebrands}}]{chumak2015magnon}
\bibinfo{author}{\bibfnamefont{A.~V.} \bibnamefont{Chumak}},
  \bibinfo{author}{\bibfnamefont{V.~I.} \bibnamefont{Vasyuchka}},
  \bibinfo{author}{\bibfnamefont{A.~A.} \bibnamefont{Serga}}, \bibnamefont{and}
  \bibinfo{author}{\bibfnamefont{B.}~\bibnamefont{Hillebrands}},
  \bibinfo{journal}{Nat. Phys.} \textbf{\bibinfo{volume}{11}},
  \bibinfo{pages}{453} (\bibinfo{year}{2015}).

\bibitem[{\citenamefont{Schiffer et~al.}(1995)\citenamefont{Schiffer, Ramirez,
  Huse, Gammel, Yaron, Bishop, and Valentino}}]{schiffer1995frustration}
\bibinfo{author}{\bibfnamefont{P.}~\bibnamefont{Schiffer}},
  \bibinfo{author}{\bibfnamefont{A.}~\bibnamefont{Ramirez}},
  \bibinfo{author}{\bibfnamefont{D.~A.} \bibnamefont{Huse}},
  \bibinfo{author}{\bibfnamefont{P.}~\bibnamefont{Gammel}},
  \bibinfo{author}{\bibfnamefont{U.}~\bibnamefont{Yaron}},
  \bibinfo{author}{\bibfnamefont{D.}~\bibnamefont{Bishop}}, \bibnamefont{and}
  \bibinfo{author}{\bibfnamefont{A.}~\bibnamefont{Valentino}},
  \bibinfo{journal}{Phys. Rev. Lett.} \textbf{\bibinfo{volume}{74}},
  \bibinfo{pages}{2379} (\bibinfo{year}{1995}).

\bibitem[{\citenamefont{Wu et~al.}(2015{\natexlab{b}})\citenamefont{Wu, Fradin,
  Hoffman, Hoffmann, and Bhattacharya}}]{wu2015spin}
\bibinfo{author}{\bibfnamefont{S.~M.} \bibnamefont{Wu}},
  \bibinfo{author}{\bibfnamefont{F.~Y.} \bibnamefont{Fradin}},
  \bibinfo{author}{\bibfnamefont{J.}~\bibnamefont{Hoffman}},
  \bibinfo{author}{\bibfnamefont{A.}~\bibnamefont{Hoffmann}}, \bibnamefont{and}
  \bibinfo{author}{\bibfnamefont{A.}~\bibnamefont{Bhattacharya}},
  \bibinfo{journal}{J. Appl. Phys.} \textbf{\bibinfo{volume}{117}},
  \bibinfo{pages}{17C509} (\bibinfo{year}{2015}{\natexlab{b}}).

\bibitem[{\citenamefont{Cornelissen et~al.}(2017)\citenamefont{Cornelissen,
  Oyanagi, Kikkawa, Qiu, Kuschel, Bauer, van Wees, and
  Saitoh}}]{cornelissen2017nonlocal}
\bibinfo{author}{\bibfnamefont{L.~J.} \bibnamefont{Cornelissen}},
  \bibinfo{author}{\bibfnamefont{K.}~\bibnamefont{Oyanagi}},
  \bibinfo{author}{\bibfnamefont{T.}~\bibnamefont{Kikkawa}},
  \bibinfo{author}{\bibfnamefont{Z.}~\bibnamefont{Qiu}},
  \bibinfo{author}{\bibfnamefont{T.}~\bibnamefont{Kuschel}},
  \bibinfo{author}{\bibfnamefont{G.~E.~W.} \bibnamefont{Bauer}},
  \bibinfo{author}{\bibfnamefont{B.~J.} \bibnamefont{van Wees}},
  \bibnamefont{and} \bibinfo{author}{\bibfnamefont{E.}~\bibnamefont{Saitoh}},
  \bibinfo{journal}{Phys. Rev. B} \textbf{\bibinfo{volume}{96}},
  \bibinfo{pages}{104441} (\bibinfo{year}{2017}).

\bibitem[{\citenamefont{Oyanagi et~al.}(2020)\citenamefont{Oyanagi, Kikkawa,
  and Saitoh}}]{oyanagi2020magnetic}
\bibinfo{author}{\bibfnamefont{K.}~\bibnamefont{Oyanagi}},
  \bibinfo{author}{\bibfnamefont{T.}~\bibnamefont{Kikkawa}}, \bibnamefont{and}
  \bibinfo{author}{\bibfnamefont{E.}~\bibnamefont{Saitoh}},
  \bibinfo{journal}{AIP Adv.} \textbf{\bibinfo{volume}{10}},
  \bibinfo{pages}{015031} (\bibinfo{year}{2020}).

\bibitem[{\citenamefont{Kikkawa et~al.}(2015)\citenamefont{Kikkawa, Uchida,
  Daimon, Qiu, Shiomi, and Saitoh}}]{kikkawa2015critical}
\bibinfo{author}{\bibfnamefont{T.}~\bibnamefont{Kikkawa}},
  \bibinfo{author}{\bibfnamefont{K.}~\bibnamefont{Uchida}},
  \bibinfo{author}{\bibfnamefont{S.}~\bibnamefont{Daimon}},
  \bibinfo{author}{\bibfnamefont{Z.}~\bibnamefont{Qiu}},
  \bibinfo{author}{\bibfnamefont{Y.}~\bibnamefont{Shiomi}}, \bibnamefont{and}
  \bibinfo{author}{\bibfnamefont{E.}~\bibnamefont{Saitoh}},
  \bibinfo{journal}{Phys. Rev. B} \textbf{\bibinfo{volume}{92}},
  \bibinfo{pages}{064413} (\bibinfo{year}{2015}).

\bibitem[{\citenamefont{Jin et~al.}(2015)\citenamefont{Jin, Boona, Yang, Myers,
  and Heremans}}]{jin2015effect}
\bibinfo{author}{\bibfnamefont{H.}~\bibnamefont{Jin}},
  \bibinfo{author}{\bibfnamefont{S.~R.} \bibnamefont{Boona}},
  \bibinfo{author}{\bibfnamefont{Z.}~\bibnamefont{Yang}},
  \bibinfo{author}{\bibfnamefont{R.~C.} \bibnamefont{Myers}}, \bibnamefont{and}
  \bibinfo{author}{\bibfnamefont{J.~P.} \bibnamefont{Heremans}},
  \bibinfo{journal}{Phys. Rev. B} \textbf{\bibinfo{volume}{92}},
  \bibinfo{pages}{054436} (\bibinfo{year}{2015}).

\bibitem[{\citenamefont{Kikkawa et~al.}(2016)\citenamefont{Kikkawa, Uchida,
  Daimon, and Saitoh}}]{kikkawa2016complete}
\bibinfo{author}{\bibfnamefont{T.}~\bibnamefont{Kikkawa}},
  \bibinfo{author}{\bibfnamefont{K.-i.} \bibnamefont{Uchida}},
  \bibinfo{author}{\bibfnamefont{S.}~\bibnamefont{Daimon}}, \bibnamefont{and}
  \bibinfo{author}{\bibfnamefont{E.}~\bibnamefont{Saitoh}},
  \bibinfo{journal}{J. Phys. Soc. Jpn.} \textbf{\bibinfo{volume}{85}},
  \bibinfo{pages}{065003} (\bibinfo{year}{2016}).

\bibitem[{\citenamefont{Sagasta et~al.}(2016)\citenamefont{Sagasta, Omori,
  Isasa, Gradhand, Hueso, Niimi, Otani, and Casanova}}]{sagasta2016tuning}
\bibinfo{author}{\bibfnamefont{E.}~\bibnamefont{Sagasta}},
  \bibinfo{author}{\bibfnamefont{Y.}~\bibnamefont{Omori}},
  \bibinfo{author}{\bibfnamefont{M.}~\bibnamefont{Isasa}},
  \bibinfo{author}{\bibfnamefont{M.}~\bibnamefont{Gradhand}},
  \bibinfo{author}{\bibfnamefont{L.~E.} \bibnamefont{Hueso}},
  \bibinfo{author}{\bibfnamefont{Y.}~\bibnamefont{Niimi}},
  \bibinfo{author}{\bibfnamefont{Y.}~\bibnamefont{Otani}}, \bibnamefont{and}
  \bibinfo{author}{\bibfnamefont{F.}~\bibnamefont{Casanova}},
  \bibinfo{journal}{Phys. Rev. B} \textbf{\bibinfo{volume}{94}},
  \bibinfo{pages}{060412} (\bibinfo{year}{2016}).

\bibitem[{\citenamefont{Angeles et~al.}(2021)\citenamefont{Angeles, Sun, Ortiz,
  Shi, Li, and Wilson}}]{angeles2021interfacial}
\bibinfo{author}{\bibfnamefont{F.}~\bibnamefont{Angeles}},
  \bibinfo{author}{\bibfnamefont{Q.}~\bibnamefont{Sun}},
  \bibinfo{author}{\bibfnamefont{V.~H.} \bibnamefont{Ortiz}},
  \bibinfo{author}{\bibfnamefont{J.}~\bibnamefont{Shi}},
  \bibinfo{author}{\bibfnamefont{C.}~\bibnamefont{Li}}, \bibnamefont{and}
  \bibinfo{author}{\bibfnamefont{R.~B.} \bibnamefont{Wilson}},
  \bibinfo{journal}{Phys. Rev. Mater.} \textbf{\bibinfo{volume}{5}},
  \bibinfo{pages}{114403} (\bibinfo{year}{2021}).

\bibitem[{\citenamefont{Kent}(1993)}]{kent1993experimental}
\bibinfo{author}{\bibfnamefont{A.}~\bibnamefont{Kent}},
  \emph{\bibinfo{title}{Experimental low-temperature physics}}
  (\bibinfo{publisher}{Macmillan International Higher Education},
  \bibinfo{year}{1993}).

\bibitem[{\citenamefont{Iguchi et~al.}(2017)\citenamefont{Iguchi, Uchida,
  Daimon, and Saitoh}}]{iguchi2017concomitant}
\bibinfo{author}{\bibfnamefont{R.}~\bibnamefont{Iguchi}},
  \bibinfo{author}{\bibfnamefont{K.-i.} \bibnamefont{Uchida}},
  \bibinfo{author}{\bibfnamefont{S.}~\bibnamefont{Daimon}}, \bibnamefont{and}
  \bibinfo{author}{\bibfnamefont{E.}~\bibnamefont{Saitoh}},
  \bibinfo{journal}{Phys. Rev. B} \textbf{\bibinfo{volume}{95}},
  \bibinfo{pages}{174401} (\bibinfo{year}{2017}).

\bibitem[{\citenamefont{Bauer et~al.}(2012)\citenamefont{Bauer, Saitoh, and van
  Wees}}]{bauer2012spin}
\bibinfo{author}{\bibfnamefont{G.~E.~W.} \bibnamefont{Bauer}},
  \bibinfo{author}{\bibfnamefont{E.}~\bibnamefont{Saitoh}}, \bibnamefont{and}
  \bibinfo{author}{\bibfnamefont{B.~J.} \bibnamefont{van Wees}},
  \bibinfo{journal}{Nat. Mater.} \textbf{\bibinfo{volume}{11}},
  \bibinfo{pages}{391} (\bibinfo{year}{2012}).

\end{thebibliography}
\end{document}